\documentstyle[aps,multicol]{revtex}

\begin{document}
\title{External losses in photoemission from strongly correlated quasi
two-dimensional solids}
\author{L. Hedin (1,2) and J. D. Lee (3,2)}
\address{1. Dept. of Physics, University of Lund, 
S\"{o}lvegatan 14A, 22362 Lund,\\
Sweden\\
2. MPI-FKF, Heisenbergstrasse 1, D-70569 Stuttgart, Germany\\
3. Dept. of Physics and Dept. of complexity science and\\
engineering,\thinspace University of Tokyo, Bunkyo ku, Tokyo 113,\\
Japan}
\date{\today}
\maketitle

\begin{abstract}
New expressions are derived for photoemission, which allow experimental
electron energy loss data to be used for estimating losses in photoemission.
The derivation builds on new results for dielectric response and mean free
paths of strongly correlated systems of two dimensional layers. Numerical
evaluations are made for $Bi_{2}Sr_{2}CaCu_{2}O_{8\text{ }}$ (Bi2212) by
using a parametrized loss function. The mean free path for Bi2212 is
calculated and found to be substantially larger than obtained by Norman et
al in a recent paper. The photocurrent is expressed as the convolution of
the intrinsic approximation for the current from a specific 2D layer with an
effective loss function. This effective loss function is the same as the
photocurrent from a core level stripped of the dipole matrix elements. The
observed current is the sum of such currents from the first few layers. The
correlation within one layer is considered as a purely two-dimensional (2D)
problem separate from the embedding three dimensional (3D) environment. When
the contribution to the dielectric response from electrons moving in 3D is
taken as diagonal in ${\bf q}$ space, its effect is just to replace bare
Coulomb potentials in the (3D) coupling between the 2D layers with
dynamically screened ones. The photo electron from a specific $CuO$ layer is
found to excite low energy acoustic plasmon modes due to the coupling
between the $CuO$\ layers. These modes give rise to an asymmetric power law
broadening of the photo current an isolated two dimensional layer would have
given. We define an asymmetry index where a contribution from a Luttinger
lineshape is additive to the contribution from our broadening function.
Already the loss effect considered here gives broadening comparable to what
is observed experimentally. Our theory is not related to the loss mechanism
recently discussed by Joynt et al, which adds additional broadening beyond
what we calculate. A superconductor with a gapped loss function is predicted
to have a peak-dip-hump lineshape similar to what has been observed, and
with the same qualitative behavior as predicted in the recent work by
Campuzano et al.
\end{abstract}

\begin{multicols}{2}

\narrowtext

\section{Introduction}

Photoemission spectroscopy (PES) is an important tool to understand the
electronic structure of strongly correlated quasi two-dimensional systems
like high $T_{c}$ superconductors. Most theoretical work concentrates on two
dimensional model systems, and when the theoretical results are compared
with PES the three-dimensionality of the actual experimental samples is only
schematically, if at all, taken into account. Further almost all discussions
are based on the sudden approximation (SA), and do not consider extrinsic
losses and interference effects. For recent work on strongly correlated
systems beyond SA we refer to Refs. \cite{Lee99}, \cite{Lee00},\cite{Joynt99}%
, \cite{Norman99}.

We define SA as the bulk one-electron spectral function augmented with
dipole matrix elements. This approximation is exact in the high energy limit
for isolated systems like atoms and molecules. For solids, where the
electrons come from a surface region and the mean free path is an important
feature, SA is never valid, not even at high energies. Here the correct high
energy limit is a convolution of the sudden approximation and the loss
function (SA*LF). SA is particularly valuable when we only look for peak
positions such as quasi-particle energies (e g for bandstructure mapping).
There are indications that also quasi-particle lineshapes are well
represented. \cite{Hedin98} When it comes to spectral properties over a more
extended energy region, which is important for e g strongly correlated
systems, SA can no longer be relied on. For core level photoemission from
weakly correlated systems like metals and valence semiconductors SA*LF
correctly describes the satellite intensities only in the keV region, while
the asymmetric quasi-particle line shape (in metals) is given correctly by
SA already at low energies \cite{Hedin98}. For {\it localized} strongly
correlated systems SA is reached rather quickly, say at 5-10 eV above
threshold. \cite{Lee99}

We analyze the three dimensional dielectric response of a stack of strongly
correlated 2D sheets in the $\left( x,y\right) $ plane, embedded in a 3D
background. We then assume, as expressed in Eq. \ref{epsbulk}, that the
response to the total electrostatic potential is given by a sum of a 3D part
and a 2D part. With the 3D part depending only on the coordinate difference
in 3D, and the 2D part on the difference in 2D, the dielectric function is
obtained on a closed form. This closed form allows us to find an approximate
relation between the electron energy loss function and the dynamically
screened potential $W$. The relation is only approximate since energy loss
is related to the diagonal part (in ${\bf q}$-space) of the dielectric
function, while we need the non-diagonal $%
\mathop{\rm Im}%
W\left( z,z^{\prime },{\bf Q},\omega \right) $ (or equivalently $%
\mathop{\rm Im}%
W\left( q_{z},q_{z}^{\prime },{\bf Q},\omega \right) $) for the loss problem
in photoemission. In PES we need to know 
${\rm Im}W(z,z^{\prime},{\bf Q},\omega)$ in the presence of a surface, while
the loss data are obtained from a bulk sample. This calls for additional
approximations.
Our numerical evaluations concern Bi2212 and are based on
a parametrization of the loss function given by Norman et al \cite{Norman99}%
. We however include dispersion in the dielectric function, which makes the
mean free path much longer. We use atomic units with $\left| e\right| =\hbar
=m=1$, and thus e g energies are in Hartrees (27.2 eV) and lengths in Bohr
radii (0.529 A).

\section{Mean free path.}

For the interpretation of photoemission from the cuprates the value of the
mean free path at energies of about 20 eV, where the experiments usually are
done, is very important. In a recent paper by Norman et al \cite{Norman99}
very short values of the order 2-3 A were obtained for Bi2212. Norman et al
however neglected the ${\bf q}$ dependence in the loss function. In the
electron gas case neglect of dispersion makes the mean free path about half
the value with dispersion. When we introduce dispersion for Bi2212 we find
an even larger effect on the mean free path.

Norman et al used a parametrization of the energy loss data on $%
Bi_{2}Sr_{2}CaCu_{2}O_{8\text{ }}$ (Bi2212) obtained by N\"{u}cker et al 
\cite{Nucker89}, 
\begin{equation}
\mathop{\rm Im}%
\frac{-1}{\varepsilon \left( \omega \right) }=\sum_{i=1}^{3}c_{i}\frac{%
\omega \Gamma _{i}\omega _{i}^{2}}{\left( \omega ^{2}-\omega _{i}^{2}\right)
^{2}+\omega ^{2}\Gamma _{i}^{2}}  \label{ImepsilonBi}
\end{equation}
with parameters (energies in eV) given below 
\[
\begin{array}{cccc}
i & c_{i} & \omega_{i} & \Gamma_{i} \\
1 & 0.164 & 1.1 & 0.7 \\
2 & 0.476 & 18.5 & 13.6 \\
3 & 0.345 & 32.8 & 17.0 
\end{array} 
\] 
The first peak at about $1\;eV$ is associated with 2D plasmon excitations,
while the large double peak comes from essentially 3D excitations since it
is similar to what is observed in Cu metal (c f Ref. \cite{Norman99}). The
linear rise for small $\omega $ comes from acoustic plasmons (due to the
coupling of the 2D plasmons in the different layers), and also to some
extent from electron-hole excitations. Phonons and other low energy 
excitations cannot be seen in N\"{u}cker et al's\cite{Nucker89} 
data since the broadening is too large (150 meV).

For an electron gas we have the well-known relation between the mean free
path $\lambda _{3D}\left( \varepsilon _{k}\right) $ and the inverse
dielectric function $%
\mathop{\rm Im}%
\varepsilon ^{-1}\left( q,\omega \right) $,

\begin{equation}
\frac{1}{\lambda _{3D}\left( {\bf \varepsilon }_{k}\right) }=\frac{2}{\pi
k^{2}}\int_{0}^{\infty }\frac{dq}{q}\int_{0}^{\omega _{\max }}%
\mathop{\rm Im}%
\left[ \frac{-1}{\varepsilon \left( q{\bf ,}\omega \right) }\right] d\omega ,
\label{invmfp3D}
\end{equation}
with 
\[
{\bf \varepsilon }_{k}=k^{2}/2,\;\omega _{\max }=\min \left(
kq-q^{2}/2,k^{2}/2-k_{F}^{2}/2\right) .
\]
In a solid at lower energies we should use Bloch functions and not plane
waves for the scattered electron. However calculations by Campillo et al 
\cite{Campillo99} show that for copper use of plane waves but with a full
bandstructure dielectric function is a reasonable approximation. In our
calculations we use Eq. \ref{invmfp3D} for the two last terms in the Norman
et al parametrization. Following Ritchie and Howie \cite{Ritchie77} 
and many other authors (cf eg ref.\cite{Tougaard84}) we
introduce dispersion by replacing $\omega _{i}$ and $c_{i}$ in Eq. \ref
{ImepsilonBi} by $\omega _{i}(q)$ and $c_{i}(q)$, 
\[
\omega _{i}(q)=\omega _{i}+\frac{q^{2}}{2},\;c_{i}(q)=\frac{c_{i}\omega
_{i}^{2}}{\omega _{i}^{2}(q)},\;i=2,3.
\]
We have put $k_{F}=0$ for simplicity, which gives a slight underestimate of
the mean free path.

The expression for the mean free path in a layered material is 
\begin{eqnarray*}
\frac{1}{\lambda _{2D}\left( {\bf k}\right) }&=&\frac{1}{\pi ^{2}k}%
\int_{-\infty }^{\infty }dq_{z}\int_{0}^{\infty }\frac{QdQ}{q_{z}^{2}+Q^{2}}%
\int_{0}^{2\pi }d\phi \theta \left( \varepsilon _{{\bf k}-{\bf q}}-\mu
\right) 
\\ \nonumber
& &\times
\mathop{\rm Im}%
\frac{-1}{\varepsilon \left( Q;\omega \right) },
\end{eqnarray*}
where $k_{z}$ and ${\bf K}$ etc are components perpendicular and parallel to
the layers, ${\bf k}=(k_{z},{\bf K})$, $k=\left| {\bf k}\right| $, $K=\left| 
{\bf K}\right| $ etc, and 
\[
\omega =\varepsilon _{{\bf k}}-\varepsilon _{{\bf k}-{\bf q}%
}=k_{z}q_{z}-q_{z}^{2}/2-Q^{2}/2+KQ\cos \phi .
\]
For simplicity we have taken free electron energies. Further considering
propagation perpendicular to the layers we have $K=0$, and no dependence on
the angle $\phi $ between ${\bf K}$ and ${\bf Q}$, 
\begin{eqnarray}\label{invmfp2D}
\frac{1}{\lambda _{2D}\left( k_{z}\right) }&=&\frac{2}{\pi k}%
\int_{0}^{k}dQ\int_{q_{\min }}^{q_{\max }}\frac{Qdq_{z}}{q_{z}^{2}+Q^{2}}%
\\ \nonumber
& &\times
\mathop{\rm Im}%
\frac{-1}{\varepsilon \left( Q;k_{z}q_{z}-q_{z}^{2}/2-Q^{2}/2\right) },
\end{eqnarray}
\[
q_{\min }=k-\sqrt{k^{2}-Q^{2}},\;q_{\max }=k+\sqrt{k^{2}-Q^{2}}.
\]
In ref.\cite{Nucker91} there is a detailed discussion of the 1 eV feature
(the first peak). It has a ${\bf Q}^2$ dispersion which is a signature of
coupled particle-holes (plasmons). They also estimate the coefficient
theoretically with reasonable results. 
From N\"{u}cker et al \cite{Nucker89} it is clear that the first peak,
besides dispersing as $Q^{2}$, quickly broadens when $Q$ increases. For the
first term in Eq. \ref{ImepsilonBi} we use 
\[
\omega _{1}(Q)=\omega _{1}+\alpha Q^{2},\;\Gamma _{1}\left( Q\right) =\Gamma
_{1}\left( 1+\frac{Q^{2}}{Q_{0}^{2}}\right), 
\]
\[
\;c_{1}\left( Q\right) =\frac{%
c_{1}\omega _{1}^{2}}{\omega _{1}^{2}(Q)},
Q_{0}=0.13\;au\text{, }\alpha
=0.6.
\]
We have put $\mu =0$. With a finite $\mu $, and thus a finite $k_{F}$ we
should replace $\sqrt{k^{2}-Q^{2}}$ with $\sqrt{k_{F}^{2}-Q^{2}}$ for $%
Q<k_{F}$ in the limits above. Such a replacement makes $1/\lambda _{2D}$
smaller and our approximation thus again slightly underestimates the mean
free path.

\begin{figure}
\vspace*{6.cm}
\includegraphics{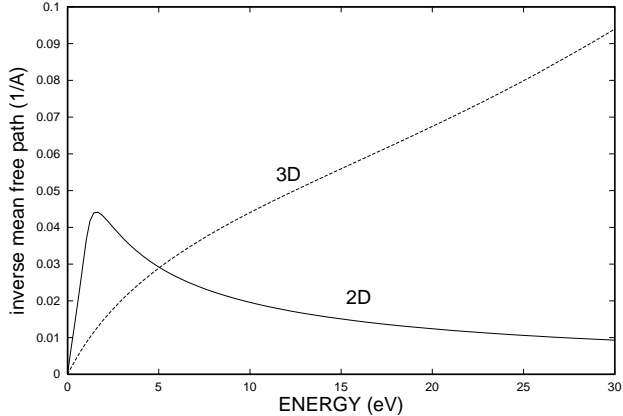}
\caption{The contributions to the inverse mean free path $1/\lambda $ from 2D
(full drawn) and 3D (dashed) terms in the case of $Bi2212$.
}
\end{figure}

\begin{figure}
\vspace*{6.5cm}
\includegraphics{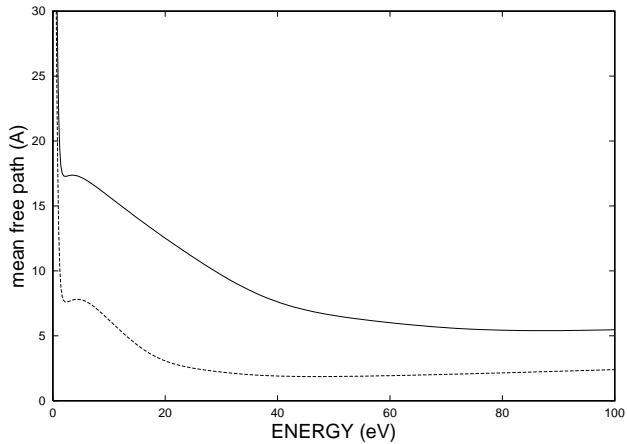}
\caption{The full drawn curve gives the mean free path $\lambda $ from our
calculations which include dispersion in the dielectric constant, and the
dashed curve the results without dispersion given by Norman et al [4].
}
\end{figure}

In Fig. 1 we show results for the inverse mean free paths for the 2D and 3D
contributions. It is remarkable that the 2D effects, while peaked at 1.5 eV,
extend quite far, out to some 30 eV. The 3D contribution starts dominating
at about 10 eV. The mean free path $\lambda $, given by $1/\lambda
=1/\lambda _{2D}+1/\lambda _{3D}$, is shown in Fig. 2. The maximum in $%
1/\lambda _{3D}$ is reached at about 100 eV, where the mean free path has
its minimum of some 5 A. It seems to be a universal feature that the minimum
mean free path is about 5 A at an energy about 3-4 times the energy where
the loss processes are strongest, as can be seen from tabulations of loss
functions \cite{Hagemann75} and mean free paths. \cite{Powell99} \ The
qualitative behavior of the 2D and 3D contributions in Fig. 1 are similar to
what has been obtained in RPA calculations for the layered electron gas. 
\cite{Hawrylak88} 
We remark that Norman et al besides the inverse mean free path also
calculated the background in PES in the traditional way from the extrinsic
losses only, following the common convention to take the background as zero
at the bottom of the main band. If we do a similar background calculation
the results are very close to the Norman et al's since in such a calculation
only the shape and not the strength of the loss function enters. We
emphasize that in our treatment of PES later in this paper we include
besides the extrinsic losses also the intrinsic ones and the interference
terms, which gives a radically different result for small energy losses.

\section{Photoemission}

We are interested in photoemission from the $CuO$ layers. The layers are
regarded as localized systems embedded in a 3D environment. The crystal
surface is taken to be parallel with the layers and in the $\left(
x,y\right) $ plane. We take the electrons in the 2D layers as separate from
the other electrons, and write the state vectors for the initial and final
states as 
\begin{equation}
\left| N_{B}\right\rangle \left| N_{2D}\right\rangle ,\;\left| N_{B}^{\ast
},s_{1}\right\rangle \left| N_{2D}-1,s_{2}\right\rangle \left| {\bf k}%
\right\rangle .  \label{eq:states1}
\end{equation}
Here $\left| N_{2D}\right\rangle $ is the state vector for the electrons in
one particular layer at a distance $z_{0}$ ($z_{0}>0$) from the surface (the
one from which the photoelectron comes), and $\left| N_{B}\right\rangle $
the state vector for the remaining (bulk) electrons which move in 3D. $%
\left| N_{2D}-1,s_{2}\right\rangle $ is an excited state $s_{2}$ of the
particular layer, and $\left| N_{B}^{\ast },s_{1}\right\rangle $ an excited
state $s_{1}$ of the bulk electrons. The star indicates that these electrons
move in the presence of a localized hole at ${\bf r}=\left( {\bf 0}%
,z_{0}\right) $, and thus is an eigenfunction of a different Hamiltonian
than that for $\left| N_{B}\right\rangle $. Finally $\left| {\bf k}%
\right\rangle $ is the photo electron state. One may argue that the hole
should be extended over the 2D layer rather than sit in one point. However
even in weakly correlated solids correlation effects give rise to satellites
corresponding rather to the removal of the electron from a point than from
an extended region. \cite{AHK96} For the strongly correlated systems
considered here the band is quite narrow and thus the atomic functions
building the Bloch functions have small overlaps, which makes our
approximation of a localized hole even better. We consider processes when
the photo electron energy is high enough that we have reached the sudden
limit as far as the excited layer is concerned (about 10 eV according to
Ref. \cite{Lee99}).

The expression for the PES\ transition amplitude then becomes \cite{Hedin99} 
\begin{eqnarray}\label{eq:tau1}
\tau \left( {\bf k},s_{1},s_{2}\right)&=&\sum_{i}\langle {\bf k}|
\langle N_{B}^{\ast },s_{1}| \langle N_{2D}-1,s_{2}|
\\ \nonumber
& &\times
\left[ 1+V\frac{1}{E-H}\right] c_{i}| N_{2D}\rangle |
N_{B}\rangle \Delta | i\rangle ,  
\end{eqnarray}
where $\left| i\right\rangle $ and $\left| {\bf k}\right\rangle $ are
one-electron states. The state $\left| i\right\rangle =\left| {\bf K}%
_{i}\right\rangle \left| \phi _{0}\right\rangle $ is the product of a 2D
Bloch state $\left| {\bf K}_{i}\right\rangle $ with momentum ${\bf K}_{i}$,
and the bound state $\phi _{0}\left( z\right) $ for the motion in the $z$
direction, which only will enter as $\left| \phi _{0}\left( z\right) \right|
^{2}=w\left( z\right) $. The operator $c_{i}$ destroys an electron in state
''$i$'', and $c_{i}\left| N_{2D}\right\rangle $ is regarded as a localized
state concerning its influence on the 3D states. The optical transition
operator is $\Delta $, and $V$ is the potential for the interaction between
the photo electron and the solid, the potential that causes external losses.
Since $V$ contains both operators acting on the photo electron and on the
solid the expectation value $\left\langle N_{B}^{\ast },s_{1}\left|
..V..\right| N_{B}\right\rangle $ is a one-electron operator acting on photo
electron states. The state $\Delta \left| i\right\rangle $ generated by
optical excitation is considered a photo electron state. $H$ is the full
Hamiltonian including $V$, and $E$ is the total energy 
\begin{eqnarray}\label{totenergy}
E&=&E\left( N_{2D}\right) +E\left( N_{B}\right) +\omega _{phot}
\nonumber \\
 &=&E\left(
N_{B}^{\ast },s_{1}\right) +E\left( N_{2D}-1,s_{2}\right) +\varepsilon _{%
{\bf k}},  
\end{eqnarray}
where $\omega _{phot}$ is the photon energy and $\varepsilon _{{\bf k}}={\bf %
k}^{2}/2$ the photo electron energy. The photocurrent is proportional to 
\[
J_{{\bf k}}\left( \omega _{phot}\right) \equiv \sum_{s_{1}s_{2}}\left| \tau
\left( {\bf k},s_{1},s_{2}\right) \right| ^{2}\delta \left( \omega
_{phot}+E_{s_{2}}-\omega _{s_{1}}-\varepsilon _{{\bf k}}\right) , 
\]
with 
\[
\omega _{s_{1}}=E\left( N_{B}^{\ast },s_{1}\right) -E\left( N_{B}\right),
\]
\[
E_{s_{2}}=E\left( N_{2D}\right) -E\left( N_{2D}-1,s_{2}\right) . 
\]

For the Hamiltonian $H$ we take, 
\[
H=H_{2D}+H_{QB}+h+V,
\]
where $H_{2D}$ describes the pertinent 2D layer, $H_{QB}$ the 3D electrons
in a quasi-boson representation, $h$ (a one-electron operator) the photo
electron, and $V$ the interaction between the photo electron and the 3D
system (the interaction with the 2D system is neglected since we assume the
sudden limit to apply here). Explicitly we have, 
\[
H_{QB}=\sum_{s}\omega _{s}a_{s}^{\dagger
}a_{s}-V_{h}P_{h},\;V_{h}=\sum_{s}V_{h}^{s}\left( a_{s}+a_{s}^{\dagger
}\right) ,
\]
\[
V=\sum_{s{\bf kk}^{\prime }}V_{{\bf kk}^{\prime }}^{s}c_{{\bf k}}^{\dagger
}c_{{\bf k}^{\prime }}\left( a_{s}+a_{s}^{\dagger }\right) ,\;V_{{\bf kk}%
^{\prime }}^{s}=\left\langle {\bf k}\left| V^{s}\right| {\bf k}^{\prime
}\right\rangle,
\]
\[
V_{h}^{s}=V^{s}\left( {\bf 0},z_{0}\right).
\]
$P_{h}$ is a projection operator which gives 1 for states with a hole in the
2D system, and 0 otherwise, and $V_{h}$ is the potential from the hole in
the 2D system. The functions $V^{s}\left( {\bf r}\right) $ are fluctuation
potentials, discussed at length in Refs. \cite{Hedin98} and \cite{Hedin99}.

Say that we somehow can calculate the photo current $J_{{\bf k}}^{2D}\left(
z_{0},\omega _{phot}\right) $ from one isolated two-dimensional layer at a
distance $z_{0}$ from the surface, and want to estimate the current from
this layer when a set of such layers together with additional electrons of
3D character form a three dimensional crystal. We have to account for the
shake up in the 3D surrounding of the layer as well as the losses the photo
electron can have on its way out to the surface. In Appendix A we show that
the photocurrent then can be written as a convolution between the 2D current 
$J_{{\bf k}}^{2D}\left( z_{0},\omega \right) $ and an effective broadening
function $P_{{\bf k}}\left( z_{0},\omega \right) $ (Eqs \ref{Jkfinal} and 
\ref{Pkfinal}), 
\begin{equation}
J_{{\bf k}}\left( z_{0},\omega _{phot}\right) =\int J_{{\bf k}}^{2D}\left(
z_{0},\omega ^{\prime }\right) P_{{\bf k}}\left( z_{0},\omega _{phot}-\omega
^{\prime }\right) d\omega ^{\prime }.  \label{mainPES}
\end{equation}
A delta function peak $\delta \left( \omega -\varepsilon _{0}-\varepsilon _{%
{\bf k}}\right) $ in $J_{{\bf k}}^{2D}\left( z_{0},\omega \right) $\ will
hence give a contribution $P_{{\bf k}}\left( z_{0},\omega
_{phot}-\varepsilon _{0}-\varepsilon _{{\bf k}}\right) $ to the
photocurrent. In core electron photoemission we have a similar expression
with $J_{{\bf k}}^{2D}\left( z_{0},\omega \right) $ replaced by the
expression for the current from a core level in an isolated ion (essentially
a delta-function).

The common EDC experiment gives the current for fixed photon energy $\omega
_{phot}$ as a function of the electron energy $\varepsilon _{{\bf k}}$ for a
given direction of ${\bf k}$ (or ${\bf K}$, where ${\bf k=}k_{z}\widehat{z}%
{\bf +K}$). We are thus interested in $P_{{\bf k}}\left( z_{0},\omega
_{phot}-\varepsilon _{0}-\varepsilon _{{\bf k}}\right) $ as function of $%
\varepsilon _{{\bf k}}$ in the range $\omega _{phot}-\varepsilon
_{0}>\varepsilon _{{\bf k}}>0$. Since $P_{{\bf k}}\left( z_{0},\omega
\right) $ varies fairly slowly with ${\bf k}$ for fixed $\omega $, $P_{{\bf k%
}}\left( z_{0},\omega \right) $ as function of $\omega $ for fixed ${\bf k}$
describes the photo emission (in a limited energy range). We will mainly
discuss the properties of $P_{{\bf k}}\left( z_{0},\omega \right) $ as
function of $\omega $ for fixed ${\bf k}$.

The effective broadening function to second order in $V^{s}$ is found to be
(Appendix A), 
\begin{equation}
P_{{\bf k}}\left( z_{0},\omega \right) =e^{-z_{0}/\lambda -a}\left[ \delta
\left( \omega \right) +\frac{\alpha \left( {\bf k},z_{0};\omega \right) }{%
\omega }\right] ,  \label{Pk}
\end{equation}
where 
\begin{eqnarray}\label{alpha1}
\frac{\alpha \left( {\bf k},z_{0};\omega \right) }{\omega }&=&\sum_{s}\left|
\int_{0}^{z_{0}}f\left( {\bf k},{\bf Q},\omega ,z_{0};z\right) V\left( q_{z},%
{\bf Q,}z\right) dz\right| ^{2}
\nonumber \\
& &\times\delta \left( \omega -\omega _{s}\right) ,\;
\end{eqnarray}
\begin{equation}
f\left( {\bf k},{\bf Q},\omega ,z_{0};z\right) =-\frac{\delta \left(
z-z_{0}\right) }{\omega }+\frac{e^{i\left( \kappa -\widetilde{k}_{z}\right)
z_{0}}}{i\kappa }e^{i\widetilde{k}_{z}z}e^{-i\kappa z}.  \label{f(z)}
\end{equation}
This expression is the same as in Eqs. 26, 27 for the core electron current
in Ref. \cite{Hedin98}. The function $V\left( q_{z},{\bf Q,}z\right) $ in
Eq. \ref{alpha1} is the fluctuation potential giving the coupling between
the photo electron and a density fluctuation $s=\left( q_{z},{\bf Q}\right) $
with energy $\omega _{s}$. In $f\left( z\right) $ the first term gives the
intrinsic or shake up contribution to the amplitude, while the second term
gives the contribution from losses when the electron propagates from the
layer at $z_{0}>0$ to the surface at $z=0$. The quantities $\kappa $ and $%
\widetilde{k}_{z}$ are the (complex) momenta in the $z$ direction of the
photo electron when inside the solid before and after it excited the density
fluctuation $s$ having parallel momentum ${\bf Q}$. The photo electron
momentum outside the solid is $k_{z}\widehat{z}{\bf +K}$ and its energy $%
\varepsilon _{{\bf k}}=\left( k_{z}^{2}+{\bf K}^{2}\right) /2$. Further $%
V_{0}$ is the (negative) inner potential, and $\Gamma _{1}$ and $\Gamma _{2}$
are the dampings before and after emitting the excitation $s$. It is easy to
derive expressions where the plane waves $e^{ikz}$ and $e^{-i\kappa z}$ are
replaced by (damped) Bloch functions, and doable to find expressions where
the bandstructure also is present in the lateral motion.

Since $P_{{\bf k}}\left( \omega \right) $ is quadratic in the fluctuation
potentials $V^{s}$, we can relate it to the dynamically screened potential $W
$. For the imaginary part of $W$ we have (c f Eq. 49 in Ref. \cite{Hedin98}%
), 
\[
\mathop{\rm Im}%
W(z,z^{\prime };{\bf R,R}^{\prime };\omega )=-\pi \sum_{s}V^{s}({\bf r}%
)V^{s}({\bf r}^{\prime })\delta (\omega -\omega _{s}).
\]
With $V^{s}({\bf r})=e^{i{\bf QR}}V\left( q_{z},{\bf Q,}z\right) $%
\begin{eqnarray*}
\mathop{\rm Im}%
W(z,z^{\prime };{\bf Q};\omega )&=&-\pi A\sum_{q_{z}}V\left( q_{z},{\bf Q,}%
z\right) V\left( q_{z},{\bf Q,}z^{\prime }\right) 
\\ \nonumber
& & \ \ \ \ \ \ \ \ \ \ \times \delta (\omega -\omega
_{s}),
\end{eqnarray*}
where $A$ is the normalization area of the planes. In an exact treatment the 
$V\left( q_{z},{\bf Q,}z\right) $ can be chosen real, and we see that $%
\mathop{\rm Im}%
W(z,z^{\prime };{\bf Q};\omega )$ is symmetric in $z$ and $z^{\prime }$.
Comparison with Eqs. \ref{Pk} and \ref{alpha1} shows that 
\begin{eqnarray*}
\frac{\alpha \left( {\bf k},z_{0};\omega \right) }{\omega }&=&-\frac{1}{\pi A}%
\sum_{{\bf Q}}\int_{0}^{z_{0}}f\left( {\bf k},{\bf Q},\omega ,z_{0};z\right)
\\ \nonumber
& \ \ &\times
\mathop{\rm Im}%
W(z,z^{\prime };{\bf Q};\omega )f\left( {\bf k},{\bf Q},\omega
,z_{0};z^{\prime }\right) ^{\ast }dzdz^{\prime }.
\end{eqnarray*}

To simplify the calculations we relate $%
\mathop{\rm Im}%
W$ to the measured loss function. The loss function however is connected
with losses in the bulk, and we also have to find an approximate relation
between $%
\mathop{\rm Im}%
W^{bulk}$ and $%
\mathop{\rm Im}%
W^{surf}$. This was done in Ref. \cite{Hedin98} by using the Inglesfield
simplified expression for the fluctuation potential, 
\begin{eqnarray*}
\mathop{\rm Im}%
W_{3D}^{surf}(z,z^{\prime },{\bf Q},\omega )&=&\frac{1}{2\pi }\int_{0}^{\infty
}F\left( q_{z},Q,z\right) F\left( q_{z},Q,z^{\prime }\right) 
\\ \nonumber
& &\times
\mathop{\rm Im}%
W_{3D}^{bulk}\left( q_{z},{\bf Q},\omega \right) dq_z,
\end{eqnarray*}
where 
\[
F\left( q_{z},Q,z\right) =2\left[ \cos \left( q_{z}z+\phi _{{\bf q}}\right)
-\cos \phi _{{\bf q}}e^{-Qz}\right] \theta \left( z\right) ,
\]
\[
\phi_{\bf q}=\arctan \frac{q_{z}}{Q}.
\]
This means that for the strength of the coupling we keep the bulk
expression, while for the spacial part we have a bulk function (here plane
wave) which is modified to be zero at the surface. The relation to the loss
function is 
\[
\mathop{\rm Im}%
W_{3D}^{bulk}\left( q_{z},{\bf Q},\omega \right) =v\left( q_{z},Q\right) 
\mathop{\rm Im}%
\frac{-1}{\varepsilon ^{3D}\left( q_{z};{\bf Q};\omega \right) }.
\]

For 2D excitations we can do a similar modification of the bulk fluctuation
potential to make it zero at the surface. When we take $w\left( q_{z}\right)
=1$ we have (see Appendix B) 
\begin{eqnarray*}
&&
\mathop{\rm Im}%
W_{2D}^{surf}(z,z^{\prime };{\bf Q};\omega )
\\ \nonumber
&=&\frac{1}{2\pi }\int_{0}^{\pi
/c}V^{r}\left( q_{z},{\bf Q,}z-z_{0}\right) 
V^{r}\left( q_{z},{\bf Q,}%
z^{\prime }-z_{0}\right) 
\\ \nonumber
& &\times
\mathop{\rm Im}%
\frac{\chi _{0}(q_{z};{\bf Q};\omega )}{c}dq_{z}
\end{eqnarray*}
for the contribution from the layer at $z_{0}$, where $%
\mathop{\rm Im}%
\chi (q;{\bf Q};\omega )$ is related to the loss function by 
\[
\mathop{\rm Im}%
\frac{\chi _{0}(q_{z};{\bf Q};\omega )}{c}=\frac{1}{2v\left( q_{z},Q\right) }%
\mathop{\rm Im}%
\frac{-1}{\varepsilon ^{2D}\left( q_{z};{\bf Q};\omega \right) }, 
\]
and $V^{r}\left( q_{z},{\bf Q,}z\right) $ is the fluctuation potential, 
\begin{equation}
V^{r}\left( q_{z},{\bf Q,}z\right) =2%
\mathop{\rm Re}%
\left[ V^{p}\left( q_{z},Q;z\right) \exp \left( -iq_{z}z+i\phi \left(
z_{0}\right) \right) \right] .  \label{Vfluct1}
\end{equation}
Here $V^{p}\left( q_{z},Q;z\right) $ is a well-known periodic potential ($%
V^{p}=\exp \left( iq_{z}z\right) V$, with $V$ defined in Eq. \ref{Vfluct3}) 
\begin{eqnarray} \label{Vfluct2}
V^{p}\left( q_{z},Q;z\right)&=&\sum_{G}v\left( q_{z}+G,Q\right) w\left(
q_{z}+G\right) e^{-iGz}
\\ \nonumber
&=&\frac{2\pi ce^{iq_{z}z}}{Q}\frac{\sinh Q\left(
c-z\right) +e^{-iq_{z}c}\sinh Qz}{\cosh Qc-\cos q_{z}c}. 
\end{eqnarray}
The explicit expression follows when the form factor $w\left( q_{z}\right) $
(c f Appendix B) is taken as $1$, and is valid only for $0<z<c$. The phase $%
\phi \left( z_{0}\right) $ in Eq. \ref{Vfluct1} is chosen to make $V^{r}$
zero at the surface, $V^{r}\left( q_{z},{\bf Q,}z_{0}\right) =0$.

\begin{figure}
\vspace*{6.5cm}
\includegraphics{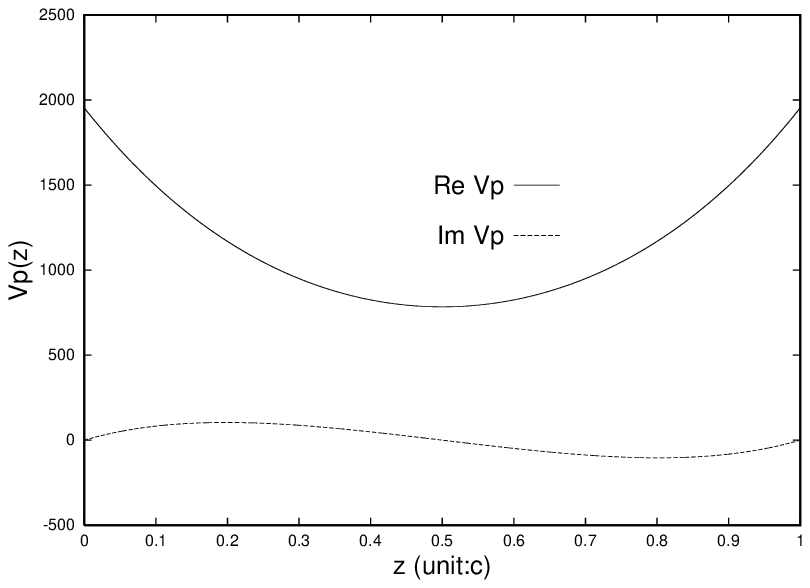}
\caption{The periodic functions $%
\mathop{\rm Re}%
V^{p\text{ }}\left( z\right) $ and $%
\mathop{\rm Im}%
V^{p}\left( z\right) $ for $z/c$ in the interval $(0,1)$, where $c$ is the
lattice constant and $V^{p}$ is defined in Eq. \ref{Vfluct2}. We have taken
some typical values, $q=0.03$ and $Q=0.1$ (for $Bi2212$ $c=29.1$ and $a=10.2$
which gives $\pi /c=0.11$ and $\pi /a=0.31$).
}
\end{figure}

In Fig. 3 we plot $%
\mathop{\rm Re}%
V^{p}\left( z\right) $ and $%
\mathop{\rm Im}%
V^{p}\left( z\right) $ for some typical values of $q_{z}$ and $Q$, and in
Fig 4 we show $V^{r}\left( z\right) $ for the same parameter values. The
sharp peak with a singular derivative in $%
\mathop{\rm Re}%
V^{p}\left( z\right) $ at $z=0$ is smoothed if we take $w\left( q\right)
\neq 1$. For a typical binding energy of 3 eV and an exponential wave
function, we have $w\left( q_{z}\right) =a^{2}/\left( a^{2}+q_{z}^{2}\right) 
$, with $a=0.9$. Typical values of $q_{z}$ and $Q$ are $\pi /c$ and $\pi /a$%
. The lattice parameters for $Bi2212$ are $c=29.1$ and $a=10.22$ which makes
$\pi /c=0.11$ and $\pi /a=0.31$.\cite{Harshman92} We can also compare with
the cut-off parameter for the collective excitations in $Bi2212$ discussed
in the section on mean free path, $Q_{0}=0.13$. Thus $a$ is substantially
larger than $q$ and $Q$, and it is hence reasonable to take $w\left(
q_{z}\right) =1$. We note that the values of $%
\mathop{\rm Re}%
V^{r}\left( z\right) $ at the first two Cu layers are substantially smaller
than the maximum value of $%
\mathop{\rm Re}%
2V^{p}\left( z\right) $. An approximation with bulk potentials cut at the
surface clearly can give very large and spurious effects unless we go to so
extremely high energies that the mean free path becomes much larger than the
lattice parameter $c$.

\begin{figure}
\vspace*{6.5cm}
\includegraphics{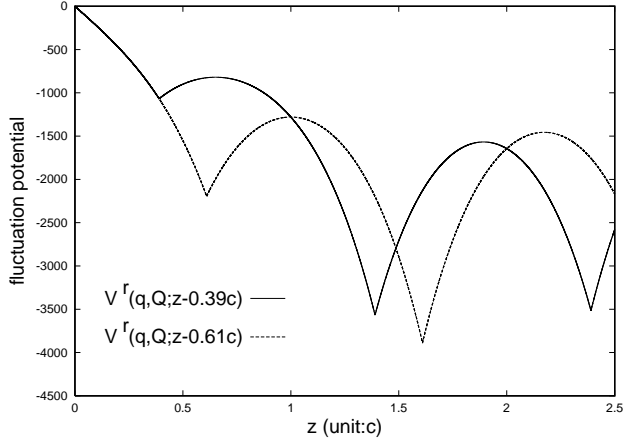}
\caption{The fluctuation potential $V^{r}(q,Q;z-z_{i})$ in Eq. \ref{Vfluct1}
for $z/c$ in the interval $(0,2.5)$ and for $z_{i}=0.39c$ and $0.61c$, the
distances from the surface of the first two $CuO$ layers. The potentials are
zero at the surface and the cusps come at $CuO$ layers. The maximum possible
value of $\left| V^{r}\right| $ is $2V^{p}\left( 0\right) $.
}
\end{figure}

When there are no low energy excitations, like for an insulator or for a
metal when the electron-hole excitations are neglected and only the plasmons
are kept, the overlap between the initial ground state and the completely
relaxed ground state in the presence of a localized hole potential is
finite. In a quasi-boson treatment we have 
\[
\left| \left\langle N_{B}^{\ast },0|N_{B},0\right\rangle \right|
^{2}=e^{-a},\;a=\sum_{s}\left| \frac{V^{s}\left( z_{0}\right) }{\omega _{s}}%
\right| ^{2}.
\]
A partial summation of the perturbation expansion in $V^{s}$ (or a cumulant
expansion) gives \cite{Hedin98}, 
\begin{equation}
P_{{\bf k}}\left( z_{0},\omega \right) =\int \frac{dt}{2\pi }
e^{-i\omega t}\exp \left( \int
\alpha \left( {\bf k},z_{0};\omega ^{\prime }\right) \frac{e^{i\omega
^{\prime }t}-1}{\omega ^{\prime }}d\omega ^{\prime }\right) .
\label{exp1}
\end{equation}
This expression correctly reproduces the edge singularity, $1/\omega
^{1-\alpha \left( {\bf k},z_{0};0\right) }$, and also gives the second order
satellite term in Eq. \ref{Pk}. In the high energy limit and the plasmon
pole approximation (the electron-hole part is then not included) it can be
shown analytically that \cite{Hedin98}, 
\begin{equation}
\int \frac{\alpha \left( {\bf k},z_{0};\omega \right) }{\omega }d\omega
=a+z_{0}/\lambda ,  \label{sumrule}
\end{equation}
and Eq. \ref{exp1} thus also gives the correct prefactor in this limit.

\begin{figure}
\vspace*{8.cm}
\includegraphics{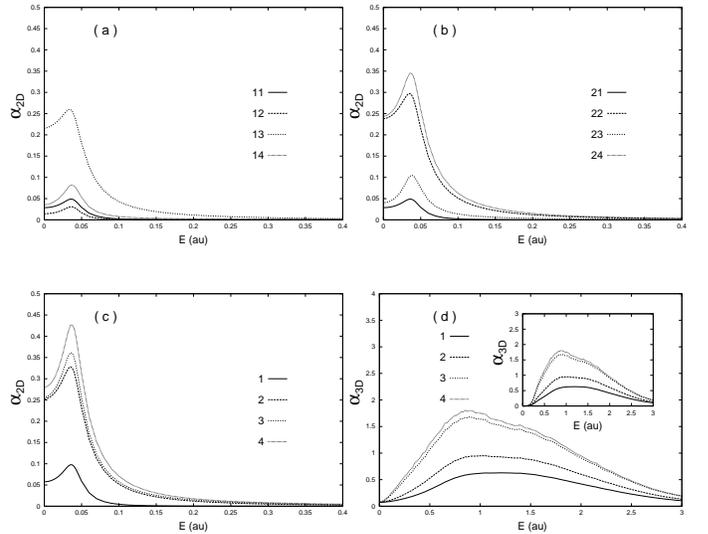}
\caption{Results for $\alpha _{2D}\left( \omega \right) $ and $\alpha
_{3D}\left( \omega \right) $ (c f Eqs. \ref{alpha2D} and \ref{alpha3D}). The
parameter values are obtained from energy loss data for $Bi2212$. The curves
in figs. 5a,b give contributions to $\alpha _{2D}\left( \omega \right) $.
The symbol mn refers to a contribution when the fluctuation potential is
centered at layer m (and m+2, m+4 etc) and the photocurrent comes from layer
n. Thus the dashed curve in fig. 5a (m=1, n=2) refers to a contribution from
the fluctuation potential centered on the layer closest to the surface when
the photocurrent comes from the second layer. Fig. 5c shows the total
contributions to $\alpha _{2D}\left( \omega \right) $ when the current comes
from layers 1 to 4, and fig. 5d the corresponding contributions to $\alpha
_{3D}\left( \omega \right) $. The curves in the inset are ad hoc adjusted to
take out the unphysical low energy part coming from a schematic
parametrization. The photon energy is $1\;au$.
}
\end{figure}

With an electron-hole continuum, Eq. \ref{sumrule} no longer can hold since
the integral diverges. We then split $\alpha \left( {\bf k},\omega
_{phot};\omega \right) $ in a 2D part from the excitations in the layers,
and a 3D part from the remaining excitations, $\alpha =\alpha _{2D}+\alpha
_{3D}$. The 3D contributions in Eq. \ref{ImepsilonBi} have been smoothly
deformed to be zero for $\omega <\omega _{th}=0.1$ since the metallic
excitations come from the layers. To give a good representation of the
experimental loss function this deformation should be compensated by a small
increase in the 2D term, but this is a minor effect which we have omitted.
Now the integral $\int_{\omega _{th}}^{\infty }d\omega \alpha _{3D}\left( 
{\bf k},\omega _{phot};\omega \right) /\omega $ converges, and we have
checked numerically that in the high energy limit 
\begin{equation}
\int_{\omega _{th}}^{\infty }\frac{\alpha _{3D}\left( {\bf k},z_{0};\omega
\right) }{\omega }d\omega =a_{3D}^{intr}+z_{0}/\lambda ,  \label{sumrule2}
\end{equation}
where $a_{3D}^{intr}$ contains only the intrinsic part, 
\[
a_{3D}^{intr}=\int_{\omega _{th}}^{\infty }\sum_{s}\left| V_{3D}^{s}\left(
z_{0}\right) \right| ^{2}\delta \left( \omega -\omega _{s}\right) \frac{%
d\omega }{\omega ^{2}}.
\]
The approach to the high energy limit is quite slow (of the order of keV),
and in our estimates for Bi2212 we adopt the expression 
\begin{eqnarray}\label{exp3}
P_{{\bf k}}\left( \omega \right)&=&e^{-z_{0}/\lambda -a_{3D}^{intr}}\int
\frac{dt}{2\pi }
e^{-i\omega t}
\\ \nonumber
& &\times
\exp \left[ \int_{0}^{\infty }\alpha _{2D}\left( {\bf k}%
,z_{0};\omega ^{\prime }\right) \frac{e^{i\omega ^{\prime }t}-1}{\omega
^{\prime }}d\omega ^{\prime }\right.
\\ \nonumber
& & \ \ \ \ \ \  \left.+\int_{\omega _{th}}^{\infty }\alpha
_{3D}\left( {\bf k},z_{0};\omega ^{\prime }\right) \frac{e^{i\omega ^{\prime
}t}}{\omega ^{\prime }}d\omega ^{\prime }\right].
\end{eqnarray}

\begin{figure}
\vspace*{7.5cm}
\includegraphics{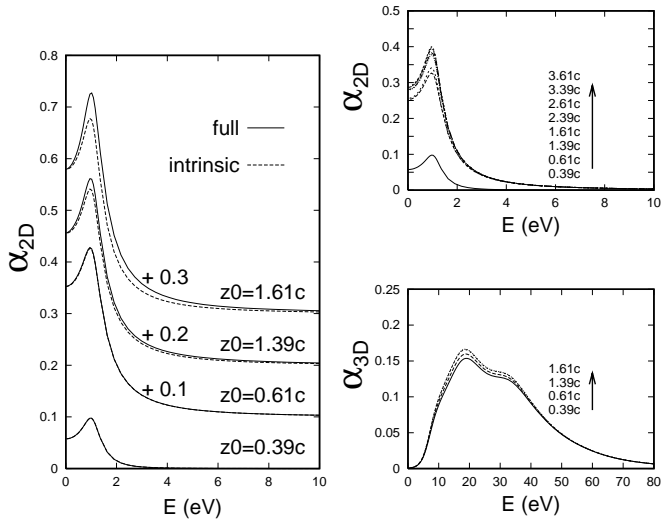}
\caption{The left part shows a comparison of $\alpha _{2D}\left( \omega
\right) $ for the pure intrinsic case with the full expression including the
extrinsic amplitude. The right part shows how the pure intrinsic
contributions to $\alpha _{2D}\left( \omega \right) $ and $\alpha
_{3D}\left( \omega \right) $ converge towards their bulk values.\ The photon
energy is $1\;au$.
}
\end{figure}

\noindent
Eq. \ref{exp3} guarantees the correct dependence on the distance $z_{0}$ to
the layer. While $\exp \left( -a_{3D}^{intr}\right) $ may not give an
accurate scale factor, this is of minor importance since it does not affect
the ratio between the threshold peak and the satellite structure. The $%
\alpha _{3D}^{intr}$ values depend only weakly on $z_{0}$, and for the first
four layers the values are $0.243,\;0.252,\;0.260,$ and $0.261$. Collecting
our results we have 
\begin{eqnarray}\label{alpha2D}
\alpha _{2D}\left( {\bf k},z_{0};\omega \right)&=&\frac{-\omega }{\pi \left(
2\pi \right) ^{3}}\int_{0}^{\pi /c}dq_{z}\int d{\bf Q}
\\ \nonumber 
&\times&\left|
\int_{0}^{z_{0}}f(z)V^{r}\left( q_{z},{\bf Q,}z\right) dz\right| ^{2}%
\mathop{\rm Im}%
\chi (q_{z},{\bf Q,}\omega ),  
\end{eqnarray}
\begin{eqnarray}\label{alpha3D}
\alpha _{3D}\left( {\bf k},z_{0};\omega \right)&=&\frac{-\omega }{\pi \left(
2\pi \right) ^{3}}\int_{0}^{\infty }dq_{z}\int d{\bf Q}
\\ \nonumber 
&\times&\left|
\int_{0}^{z_{0}}f(z)F\left( q_{z},Q,z\right) dz\right| ^{2}%
\mathop{\rm Im}%
W_{b}\left( q_{z},{\bf Q},\omega \right) .  
\end{eqnarray}

\begin{figure}
\vspace*{10.5cm}
\includegraphics{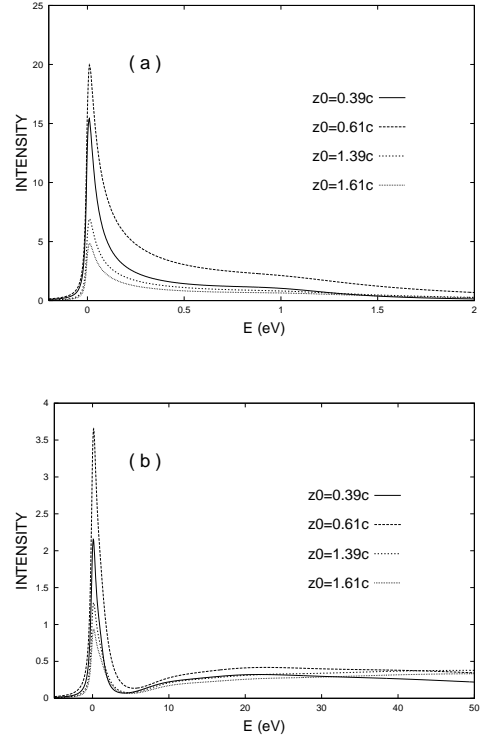}
\caption{The effective loss function $P_{{\bf k}}\left( \omega \right) $
including both 2D and 3D contributions (c f Eq. \ref{exp3}). The
contributions from the different layers are displayed separately. The curves
are convoluted with Lorentzians, in Fig. 7a with $\Gamma =10\;meV$, and in
Fig. 7b with $\Gamma =300\;meV$.
}
\end{figure}

In Fig. 5 we plot results for different contributions to the $\alpha $
functions in Eq. \ref{exp3}. The general shape of the $\alpha _{2D}\left(
\omega \right) $ functions is similar to the electron gas case with a flat
portion for small $\omega $ followed by a plasmon peak (c f Ref. \cite{AH83}%
, pp 663-667). However the magnitudes are different, $\alpha _{2D}\left(
0\right) $ is fairly large ($0.25-0.30$) compared to metals while the
plasmon peak is much smaller and broader. When we change the parametrization
to make the 3D terms start at $0.1\;au$ the $\alpha _{2D}\left( 0\right) $
values will increase by some $10\%$. The $\alpha _{2D}$ functions have only
a weak dependence on photon energy, while the $\alpha _{3D}$ curves have a
much larger dependence. All curves in Fig. 5 are for the same photon energy
of $1\;au.$

\begin{figure}
\vspace*{11.cm}
\includegraphics{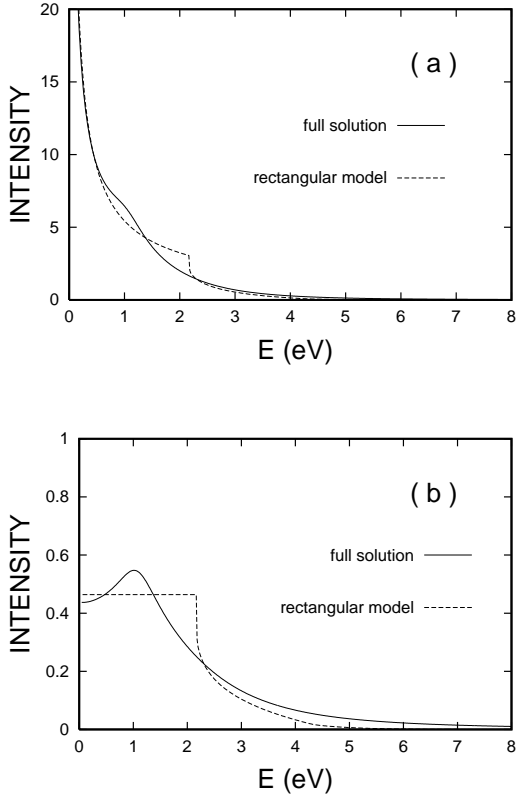}
\caption{The effective loss function $P_{{\bf k}}^{2D}\left( \omega \right) $
obtained with only the 2D contribution from the second layer. The full drawn
curve in fig 8a shows the full solution obtained with the ''2'' curve in
Fig. 5c, while the dashed curve shows the result using the rectangular
approximation for $\alpha _{2D}\left( \omega \right) $ with $\omega
_{0}=0.08 $ and $\alpha _{0}=0.255$ (see text at Eq. \ref{Pth}). In fig. 8b
we show $\omega ^{\left( 1-\alpha \left( 0\right) \right) }P_{{\bf k}}\left(
\omega \right) $. The photon energy is $1\;au$.
}
\end{figure}

The left part of Fig. 6 shows the dominance of the intrinsic contributions
to $\alpha _{2D}$. As expected the contributions from the extrinsic terms
are larger for the layers further away from the surface. The right part
shows the approach towards the bulk value of the intrinsic contributions for 
$\alpha _{2D}$ and $\alpha _{3D}$. This approach is considerably slower in
the 2D case as might be expected from the behavior of the fluctuation
potentials (c f Fig. 4). Comparing the intrinsic $\alpha _{3D}$ in Fig. 6
with the full $\alpha _{3D}$ in Fig. 5, we see that in the 3D case the
extrinsic effects dominate. The difference $\alpha _{3D}^{full}-\alpha
_{3D}^{intr}$ is roughly proportional to $z_{0}$ which follows the trend in
the high energy sum rule, Eq. \ref{sumrule2}. In Fig. 7 the contributions
from $P_{k}\left( z_{0},\omega \right) $ in Eq. \ref{exp3} from the first
four $CuO$ layers are shown. It is clear that most of the asymmetry comes
from the layers in the first unit cell. The alpha function for the first
copper layer is quite small (Fig. 6), but when the mean free path effects
are taken into account, Fig. 7 shows that the broadening contributions from
the first and second layers are comparable. Fig 7b shows an extended energy
region to illustrate the relative importance of the 2D and 3D\
contributions. The integral effect of the 3D contributions is much larger,
but the peaks in the loss function are smoothed out and the 3D contribution
is featureless. At higher energies we of course also have contributions to
the photo current from other states than the quasi 2D ones in the Copper
layers discussed in this paper.

We now give a qualitative discussion of the effective broadening function $%
P_{{\bf k}}\left( \omega \right) $ in Eq. \ref{exp3}. Since the 2D and 3D
contributions add in an exponent we can write $P_{{\bf k}}\left( \omega
\right) $ as a convolution, 
\[
P_{{\bf k}}\left( z_{0},\omega \right) =e^{-z_{0}/\lambda
-a_{3D}^{intr}}\int P_{{\bf k}}^{2D}\left( \omega -\omega ^{\prime }\right)
P_{{\bf k}}^{3D}\left( \omega ^{\prime }\right) d\omega ^{\prime }.
\]
For $P_{{\bf k}}^{3D}$ we make a Taylor expansion, and keep only the first
term, $P_{{\bf k}}^{3D}\left( \omega \right) =\delta \left( \omega \right)
+\alpha _{3D}\left( \omega \right) /\omega $. We have then omitted the
multiple quasi-boson excitations starting at $\omega =2\omega _{th}$. Since $%
P_{{\bf k}}^{2D}$ is normalized to unity, and consists of a peak that is
sharp compared to $\alpha _{3D}$, we can write 
\[
P_{{\bf k}}\left( z_{0},\omega \right) \simeq e^{-z_{0}/\lambda
-a_{3D}^{intr}}\left[ P_{{\bf k}}^{2D}\left( z_{0},\omega \right) +\frac{%
\alpha _{3D}\left( {\bf k},z_{0};\omega \right) }{\omega }\right] .
\]
To numerically evaluate $P_{{\bf k}}^{2D}\left( \omega \right) $ we used the
integral equation $\omega P_{{\bf k}}^{2D}\left( \omega \right)
=\int_{0}^{\omega }d\omega ^{\prime }\alpha _{2D}\left( \omega ^{\prime
}\right) P_{{\bf k}}^{2D}\left( \omega -\omega ^{\prime }\right) $ which is
easier then to evaluate the exponential expression in Eq. \ref{exp3}. If we
approximate $\alpha _{2D}\left( \omega \right) $ by a rectangular function, $%
\alpha _{2D}\left( \omega \right) =\alpha _{0}\theta \left( \omega
_{0}-\omega \right) $, and broaden with a Lorentzian of width $\Gamma $ $%
\left( FWHM=2\Gamma \right) $, we have for $\omega <\omega _{0}$ the
Doniach-Sunjic expression \cite{Doniach70}, 
\[
P_{{\bf k}}^{2D}\left( \omega \right) =C\left( \alpha _{0}\right) \frac{\cos %
\left[ \pi \alpha _{0}/2-\left( 1-\alpha _{0}\right) \arctan \left( \omega
/\Gamma \right) \right] }{\left( 1+\left( \omega /\Gamma \right) ^{2}\right)
^{\left( 1-\alpha _{0}\right) /2}},
\]
\begin{equation}
C\left( \alpha _{0}\right) =\frac{%
e^{-\gamma \alpha _{0}}}{\left( \alpha _{0}-1\right) !\omega _{0}^{\alpha
_{0}}\Gamma ^{1-\alpha _{0}}\sin \left[ \pi \alpha _{0}\right] }  \label{Pth}
\end{equation}
where $\gamma =0.577$ is the Euler constant. The coefficient $C\left( \alpha
_{0}\right) $ in Eq. \ref{Pth} was derived in Ref. \cite{AH83} (see Eq.
162). For $\omega >\omega _{0}$ $P_{{\bf k}}^{2D}\left( \omega \right) $
only has a weak tail with less than 10\% of the norm (for $\alpha _{0}<0.4)$%
. Let $\omega _{\max }$ be the $\omega $ value for which $P_{{\bf k}%
}^{2D}\left( \omega \right) $ has its maximum, and $\omega _{1}$ and $\omega
_{2}$ the values where it takes half its maximum value. We define an
asymmetry index $\gamma _{as}\left( \alpha _{0}\right) =$ $\left( \omega
_{2}-\omega _{\max }\right) /\left( \omega _{\max }-\omega _{1}\right) $. An
approximate expression for $\gamma _{as}$ is 
\begin{equation}
\gamma _{as}\left( \alpha _{0}\right) =\frac{\omega _{2}-\omega _{\max }}{%
\omega _{\max }-\omega _{1}}=1+0.79\alpha _{0}+14.54\alpha _{0}^{2}.
\label{gammaas}
\end{equation}

$P_{{\bf k}}^{2D}\left( \omega \right) $ is the function that broadens a $%
\delta $-function peak in $J_{{\bf k}}^{2D}\left( \omega \right) $. If $J_{%
{\bf k}}^{2D}\left( \omega \right) $ has a Doniach-Sunjic singular shape the
broadening with $P_{{\bf k}}^{2D}\left( \omega \right) $ still gives Eq.\ 
\ref{Pth} but with an $\alpha _{0}$ that is the sum of the alphas in $J_{%
{\bf k}}^{2D}$ and in $P_{{\bf k}}^{2D}\left( \omega \right) $. This is so
because the time transform of a power law singularity $\omega ^{-\left(
1-\alpha _{0}\right) }$ is $t^{-\alpha _{0}}$, and a convolution in
frequency space is a product in time space.

\begin{figure}
\vspace*{11.cm}
\includegraphics{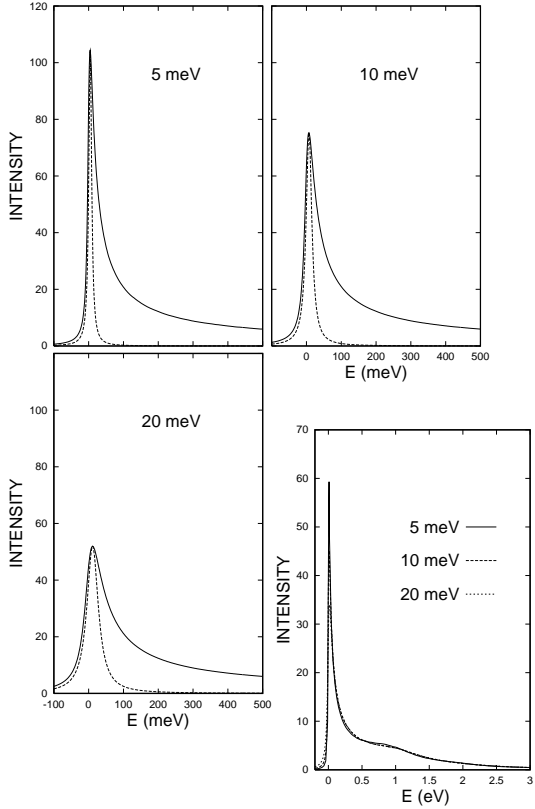}
\caption{The effective loss function $P_{{\bf k}}\left( \omega \right) $
convoluted with Lorentzians of different widths $\Gamma $ (5, 10 and 20
meV). All the 2D contributions $P_{{\bf k}}^{2D}\left( \omega \right) $ in
Eq. \ref{Pth} from the first four layers are summed weighted with $\exp
(-z_{0}/\lambda -\alpha _{3D}^{intr})$. The 3D terms are not included. In
the first three panels we have used the rectangular approximation, while in
the last panel the full evaluation from Eq. \ref{exp3} was done. Also the
Lorentzians are shown. The photon energy is $1\;au$.
}
\end{figure}

The $\alpha _{2D}$ functions in Fig. 5 show clear peaks due to the plasmon
excitations. The peaks are however not strong enough to give more than a
small bump in the $P_{{\bf k}}^{2D}$ functions. This is illustrated in Fig.
8a which shows the full $P_{{\bf k}}^{2D}$ curve and the rectangular
approximation in Eq. \ref{Pth} using $\alpha _{0}=0.255$ and $\omega
_{0}=0.08\;au=2177\;meV$. In Fig 8b the rectangular approximation is
illustrated by taking out the singularity and plotting $P_{{\bf k}%
}^{2D}\left( \omega \right) \omega ^{\left( 1-\alpha _{2D}\left( 0\right)
\right) }$. The simple rectangular approximation without plasmon peak should
be useful as a guide when other broadening effects are at work. In Fig. 9 we
show the sum for the first four layers of the 2D contributions $\exp \left(
-z_{0}/\lambda -a_{3D}^{intr}\right) P_{{\bf k}}^{2D}\left( \omega \right) $
broadened with different Lorentzians. The 3D terms are not included except
for the (all important) mean free path factor. In the first three panels
with a limited energy region (up to 500 meV) we have used the rectangular
approximation for the different $\alpha _{2D}$ contributions. In the last
panel with a larger energy range the full evaluation from Eq. \ref{exp3} was
done since it is superior to the rectangular model for energies above 0.5 eV
(see Fig. 8). The numerical accuracy at the peak is however lower in the
full calculation. Also the Lorentzians are shown to ease the estimate of the
size of the asymmetries. It is clear that we have a sizeable line asymmetry,
and also a long tail extending over several eV. The artificial step in the
rectangular approximation at about 3 eV (Fig. 8) is of little consequence
since the intensity is small at this energy. The asymmetry index is
slightly dependent on the Lorentzian broadening $\Gamma $ since we have
summed contributions from different layers with different $\alpha $ values.
The index is about $2.6$ which according to Eq. \ref{gammaas} corresponds to
an effective $\alpha $ of about $0.3$.

In the superconducting state the loss function should have a gap. We mimic
this gap by using a rectangular alpha function 
\begin{equation}
\alpha _{2D}\left( \omega \right) =\alpha _{0}\theta (\omega -\omega
_{sc})\theta \left( \omega _{0}-\omega \right) ,  \label{alpha2D1}
\end{equation}
using the same values for $\alpha _{0}$ and $\omega _{0}$ as in Fig. 8. For
the gap $\omega _{sc}$ we take $\omega _{sc}=70\;meV$. In Fig. 10 we show
the corresponding $P_{{\bf k}}^{2D}\left( \omega \right) $ broadened with a
Lorentzian of width $\Gamma =15\;meV$. Our choice of parameters is only made
to illustrate the qualitative behavior to be expected. The curve clearly
shows the peak-dip-hump lineshape found experimentally (for a recent
reference see e g \cite{Campuzano99})

\begin{figure}
\vspace*{6.cm}
\includegraphics{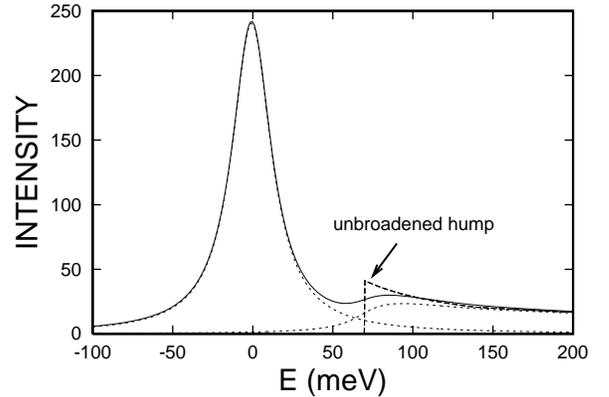}
\caption{The loss function $P_{{\bf k}}^{2D}$ for a gapped spectrum using
the simple parametrization in Eq. \ref{alpha2D1}. The Lorentzian broadening
is $\Gamma =15meV$.
}
\end{figure}

Recently it has been possible to obtain very accurate tunneling data from
Bi2212, and it is of interest to compare these data with the PES satellites
(ref.\cite{Note}), since the tunneling data also show peak-dip-hump
structures. \cite{Yurgens99} \ PES and tunneling 
are basically different spectroscopies. There can however
be qualitative similarities since in both cases the electrons couple to 3D
quasi-boson excitations like phonons, electron-hole pairs, plasmons, magnons
etc. In our treatment of PES we take the states of a particular 2D layer as
given and study the effect to low order of the sudden appearance of a hole
in the 2D system on the quasi-bosons (intrinsic excitations) as well as of
the coupling of the photoelectron leaving this layer to the quasi-bosons
(extrinsic excitations), and their interference (c f Eqs. 8-10, or
equivalently Eq. 26 in Appendix A). We found that the intrinsic
contributions dominate for small excitation energies. 

Tunneling is traditionally described by the spectral function which involves
matrix elements of the electron annihilation operator between the initial
state and the excited states. \cite{Schrieffer63},\cite{Mahan81} The
excited states consist of a 2D layer state with a hole, and some state of
the quasi-bosons in the presence of a localized hole. In lowest order
perturbation theory the probability for a final state with excited
quasi-bosons is given by the first term in Eq. 10. This means that {\it the
intrinsic contribution to PES and the tunneling currents are the same except
for} the mean free path effect in PES shown in Eq. 8, and the summation over
momenta in tunneling (giving the density of states, DOS) contra momentum
conservation from the dipole matrix element in PES. As mentioned above, we
modify this analysis valid for the normal state, by simply assuming that the
loss function should have a gap in the superconducting state.

In Bi2212 we have a van Hove singularity (VHS) at the Fermi level, which
makes the difference between DOS and momentum conservation of less
importance (there might actually even be two VHS if the two $CuO$ planes at $%
3\;A$ apart produce a significant splitting). More important is that in PES
the electrons come from a thin surface region (of the order of the mean free
path) while in tunneling they may come from an extended region which can be
hundreds of $A$, and that the coupling functions $V(z)$ are strongly $z$%
-dependent. Additionally there are two energy gaps (superconducting gap and
pseudogap), which further complicates the picture. There is thus no way that
PES and tunneling structures can be quantitatively the same, but since the
same quasi-bosons are involved, there may well be qualitative similarities
even though the coupling strengths can be quite different. It should also be
noted that we take the spectral function for the 2D system as a sharp peak
(the function often calculated by theoreticians using say a $t-J$ model), and
have no means to estimate the relative strengths of the true 2D spectral
function relative to the loss structure analyzed here. 

In our analysis we have only treated the plasmons for the simple reason that
the experimental loss data at hand did not have resolution enough to show
phonons and other low energy excitations. If such ($q=0$) data appear
showing additional quasi-bosons one is faced with introducing reasonable
dispersions, and finding reasonable extrapolations of the bulk coupling
function to account for the presence of the surface.

\section{Concluding remarks}

This paper is concerned with effects of external losses in photo emission,
and the extent to which the commonly used sudden approximation works for
strongly correlated layered materials. We have earlier found that for a
strongly correlated {\it localized} system the sudden approximation is
reached rather quickly, at about $10\;eV$ \cite{Lee99}. For a weakly
correlated system on the other hand, like an sp-metal or semiconductor, the
sudden limit is approached very slowly, on the keV scale. \cite{Hedin98} \
The slow approach is connected with strong destructive interference between
the intrinsic and extrinsic mechanisms for plasmon production. The
cancellation is particularly strong for small momentum plasmons where the
long-wave plasmons are excited by the average potential from the core hole
and photo electron, which is zero \cite{Gadzuk77}. The asymmetric lineshape
in core electron photo emission from metals is, on the other hand, hardly
affected by the external loss processes \cite{Hedin98}.

We are interested in energies where the sudden limit is reached for the
strongly correlated layer from which the photo electron comes, and derive an
expression for the photo current as a convolution of the sudden
approximation for the current from the layer with an effective loss
function, $P_{{\bf k}}\left( \omega \right) $ (Eq. \ref{mainPES}). We
assume, as far as the loss properties are concerned, that the photo electron
comes from a localized position. In our specific example, $Bi2212$, the $c$
value is $15.4\;A$ (neglecting crystallographic shear), and almost all
contributions come from the first unit cell. The two first $CuO$ layers are
at $0.39\;c$ and $0.61\;c$ from the surface (which is between two $BiO$
layers). \cite{Takahashi89}With a photon energy of $1\;au$ the maximum
electron energy (inside the solid) is $1.15\;au=31\;eV$ if we take the
bandwidth as $0.15\;au$. Our energy loss calculations give $\lambda
=17.8\;a_{0}$ (Fig. 2). The $\exp (-z_{0}/\lambda )$ factor then is $%
0.53,\;0.37,\;0.10,$ and $0.07$ for the first four $CuO$ layers. We thus
expect large photo emission contributions only from the first unit cell.

To obtain $P_{{\bf k}}\left( \omega \right) $ we use a previously developed
method based on a quasi boson\ model, where the electron-boson coupling is
given by fluctuation potentials related to the dielectric response function 
\cite{Hedin99}. We find that the energy loss function, which we take from
experimental data, can be related to the screened potential which we need to
calculate the (intrinsic and extrinsic) losses in photo emission. The
fluctuation potentials related to the electrons in the layers are universal
functions, which are easily calculated (Eqs \ref{Vfluct1} and \ref{Vfluct2}%
). They have some resemblance to a surface plasmon potential, but penetrate
the whole solid and have the Bloch wave symmetry. We use the real part (or
equivalently the imaginary part) of a phase shifted bulk potential to get a
potential which is zero at the surface, and mimics the potential we have in
a finite solid. The fluctuation potential is integrated over $z$ together
with a propagation function $f\left( z\right) $ (Eq. \ref{f(z)}) that takes
the photo electron out of the solid.\ This integral is in turn integrated
with the loss function (taken from experiment), Im$\varepsilon ^{-1}\left(
q,Q,\omega \right) $, to give functions $\alpha _{2D}\left( \omega \right) $
and $\alpha _{3D}\left( \omega \right) $ which are simply related to the
effective loss function $P_{{\bf k}}\left( \omega \right) $ (Eq. \ref{exp3}%
). When we use plane waves instead of Bloch functions in the propagation
function $f\left( z\right) $, all specific materials properties are embodied
in the loss function. The propagation function has both an intrinsic and an
extrinsic contribution which interfere.

From Eq. \ref{exp3} we see that $P_{{\bf k}}\left( \omega \right) $ is
scaled down with $z_{0}$, the distance of the layer from the surface, while
the fluctuation potentials increase with $z$. The reason for that increase
is that the boundary condition forces the fluctuation potential in the first
unit cell to be much weaker than the bulk potential (c f Figs 3 and 4). The
contributions to the $\alpha $ functions from excitations in different
layers are shown in Fig 5 for photoemission from different layers.

The mean free path is found to be considerably longer than obtained by
Norman et al \cite{Norman99}, about $12\;A$ rather than $3\;A$, at say $%
20\;eV$ (Fig. 2). Measurements by the ITR-2PP technique \cite{Nessler98}
give a lifetime of $\tau =10\;fs$ at an energy $\varepsilon =3\;eV\,$\ above
the Fermi surface. The mean free path is $\lambda =v\tau $. Converting
energy to velocity by $mv^{2}/2=\varepsilon $ gives a mean free path $\tau
=103\;A$ as compared to our result of about $17\;A$ at that energy. This is
an indication that our values rather are on the low side. It is however hard
to know what is the correct conversion between energy and velocity at such
low energies, which makes a comparison very uncertain.

From Fig. 1 we see that the 2D losses occur only for small energies, at $%
5\;eV$ the bulk losses take over. The 2D losses go to zero quite slowly,
just like the bulk losses, but on another energy scale. If we only had 2D
losses, the minimum mean free path would be long, about $20\;A$. The general
behavior of the 3D\ mean free path follows a well known pattern. The mean
free path has a minimum of about $5\;A$ at an energy of 3-4 times the energy
where the loss function has its center of gravity. We have used the Born
approximation to evaluate the mean free paths. This may seem a very crude
approximation at low energies. However the Born scattering expression with a
basis of Bloch waves and Bloch energies rather than plane waves and free
electron energies agrees with the GW approximation, which is commonly used
also at low energies. Further it was shown by Campillo et al \cite
{Campillo99} that plane waves and free electron energies was not that bad,
as long as the energies in the dielectric function are well approximated.

Our main concern is the behavior of the effective broadening function at
small energies where it is dominated by the 2D losses. The 3D contributions
set in at somewhat higher energies, and give a rather structure-less
contribution. What we here for convenience call 2D losses is of course
actually also a 3D effect since it comes form excitations of a coupled set
of 2D layers. To allow a qualitative discussion we represent the $\alpha
_{2D}$ functions by a rectangular distribution. Looking at Fig. 5 this may
seem rather crude, but Fig. 8 shows that the corresponding $P_{{\bf k}%
}\left( \omega \right) $ functions are not too different. The rectangular
distribution allows an analytic solution (Eq. \ref{Pth}) valid out to the
cut-off $\omega _{0}$ ($\omega _{0}\simeq 0.1\;au\simeq 3\;eV$). $P_{{\bf k}%
}\left( \omega \right) $ has only a fairly small tail beyond $\omega _{0}$ .
In Fig. 9 we plot the total $P_{{\bf k}}\left( \omega \right) $ function
(sum over the four first layers, properly mean free path weighted),
calculated with the rectangular approximation and broadened with Lorentzians
of different widths. We note the marked asymmetry. The asymmetry is
described by an index $\gamma _{as}$, defined in Eq. \ref{gammaas}. When $%
P_{k}\left( \omega \right) $ derives from only one (rectangular) $\alpha $%
-function $\gamma _{as}$ is a function of the singularity index $\alpha
_{0},\;\gamma _{as}\left( \alpha _{0}\right) $. The index $\gamma _{as}$ is
then independent of the amount of Lorentzian broadening $\Gamma $. If the $%
J_{2D}$ function has a power law singularity with singularity index $\alpha
_{L}$, the asymmetry index contains the sum of the two indices, $\gamma
_{as}\left( \alpha _{0}+\alpha _{L}\right) $.

In Fig. 7 we plot contributions to the loss function $P_{k}\left( \omega
\right) $ from different layers. It is interesting that the first two layers
give about the same contribution, while the contributions from the next two
are tiny. In the left part of Fig. 6 we show the importance of the intrinsic
contributions to $\alpha _{2D}$. The behavior here is thus similar to what
was found previously for metals. \cite{Hedin98} In the right part of Fig. 6
we show the approach to the bulk limit curves. This approach is very slow
for $\alpha _{2D}$ while, like in metals, it is fast for $\alpha _{3D}$. The
slow approach for $\alpha _{2D}$ of course comes from the slow approach to
the bulk limit of the 2D fluctuation potentials (Fig. 4).

In a paper by Liu, Anderson and Allen from 1991 \cite{Liu91}, they discussed
the lineshapes of $Bi_{2}Sr_{2}BaCu_{2}O_{8}$ along the $\Gamma -X$
direction obtained by Olsen et al \cite{Olson90} for $22\;eV$ photons. They
concluded that neither the Fermi liquid nor the marginal Fermi liquid
theories could fit the slow fall-off of the spectrum at higher energies. Our
results offer a possibility that the slow falloff may be due to intrinsic
creation of acoustic plasmons in a coupled set of $CuO$ layers, an effect
not present if only one $CuO$ layer is considered. This broadening is mostly
intrinsic, i e if we treat a 3D system we have an almost intrinsic effect.
However most theoretical discussions concern an isolated 2D system, compared
to which we find an appreciable extra broadening from the coupling between
the layers.

The PES spectra change strongly when we go to the superconducting state. The
main peak sharpens and a peak-dip-hump structure develops. This effect has
been interpreted as a coupling of the 2D\ state to the $\left( \pi ,\pi
\right) $ collective mode. \cite{Campuzano99} \ Here we find that this
effect also can arise from the gapping of the loss function caused by the
lack of low energy excitations in a superconductor as shown in Fig. 10.
Without a more accurate model we find it difficult to decide which is the
correct explanation, possibly it could be a combination of both mechanisms.
Since the gapping of the loss function is related to the superconducting
gap, also with our mechanism the hump will scale with the gap. It is clear
that the experimental peak-dip-hump structure rides on a background which is
not predicted by our expressions, nor by anyone else's. Our theory is
however rather schematic with its strict separation of a 2D and a 3D part,
while in reality the bands are hybridized. If we extend our approach to a
more detailed treatment of the underlying bandstructure, the background
could well be strongly changed. Such an extension represents a very large
numerical task but with the present pilot treatment we can at least start
thinking seriously about the difficult background problems in photoemission.

In recent papers Joynt et al \cite{Joynt99} discussed a broadening mechanism
due to the interaction between the photo electron when outside the solid and
the electrons in the solid. This is a different mechanism than in this
paper, which adds additional broadening. Their discussion only involved the
energy loss part and not the elastic contribution and can thus not be
directly compared to experiment. We hence find their claims regarding
pseudogaps uncertain.

It should be stressed that we cannot claim any high quantitative accuracy.
We have put in dispersion in the loss function using a crude approximation.
Since however dispersion is very important we think our predictions are
substantially better than if dispersion had been neglected. We have only
considered normal emission where the electrons come from the $\Gamma $
point, while the interesting experiments concern electrons from the Fermi
surface. However there is no reason that the effective loss function should
change qualitatively when we go away from normal emission. The behavior of
the loss function when $\omega \rightarrow 0$ has been disputed. Most
authors seem to believe the approach is linear, but there are also claims
that it should be quadratic. \cite{Bozovic90} If it were quadratic, the
corresponding $\alpha $-function would start linearly rather than with a
constant. However $\alpha \left( \omega \right) $ would have to rise very
fast to reproduce the behavior of the loss function for the (quite small)
energies where it is known to be approximately linear. Thus the pure power
law behavior of $P_{k}\left( \omega \right) $ would be lost, but Lorentzian
broadened curves would probably not differ much. Our fluctuation potentials
are obtained by phase-shifting bulk potentials to make them zero at the
surface, and define them as zero outside the solid. This procedure turned
out to be fairly good in the metallic case, where we could check with more
accurately calculated fluctuation potentials. Again this approximation is
crude, but we believe it to be fundamentally better than if we had used a
step function on the bulk potential. Since the phase of the bulk potential
is arbitrary, such a procedure would anyhow have been arbitrary. To
calculate more accurate potentials\ is a very large numerical undertaking.

One may also question the use of a bulk expression to estimate of the mean
free path at the fairly low energies that we are concerned with, after all
we found strong effects when \ modifying the fluctuation potentials for
surface effects. It does not seem easy to make a strong statement here, and
we can only refer to ''the state of the art'', that bulk mean free paths are
successfully used in LEED and also in low energy life time calculations
which are compared with time-resolved two-photon PES (TR-2PPE) experiments.
\cite{Echenique00}

\section*{acknowledgement}

We thank J.W. Allen, J.C. Campuzano, A. Fujimori, A.J. Millis, M.R. Norman, 
and Z.-X. Shen for constructive and informative comments.
One of the authors (J.D.L.) acknowledges the fellowship
from the Japan Society for the Promotion of Science.

\section{Appendix A. Derivation of the photo current expression}

We will here derive Eq. \ref{mainPES}. The 2D and 3D parts in Eq. \ref
{eq:tau1} factor, 
\[
\tau \left( {\bf k,}s_{1,}s_{2}\right) =\sum_{i}\left\langle
N_{2D}-1,s_{2}\left| c_{i}\right| N_{2D}\right\rangle \tau ^{3D}\left( {\bf %
k,}s_{1},i\right) , 
\]
where 
\begin{eqnarray*}
& &\tau ^{3D}\left( {\bf k,}s_{1},i\right)
\\ \nonumber
&\equiv& \langle {\bf k}|
\langle N_{B}^{\ast },s_{1}|
\\ \nonumber
&\times&
\left[1+V\frac{1}{E\left( N_{B}^{\ast
},s_{1}\right) +\varepsilon _{{\bf k}}-H_{QB}-h-V}\right]
|N_{B}\rangle
\Delta | i\rangle . 
\end{eqnarray*}
We note that $\left\langle N_{2D}-1,s_{2}\left| c_{i}\right|
N_{2D}\right\rangle $ is the basic part in the spectral function for the 2D
system, and that the 2D and 3D parts are entangled through the index $i$. We
have used Eq. \ref{totenergy} to eliminate the index $s_{2}$, $E-E\left(
N_{2D}-1,s_{2}\right) =E\left( N_{B}^{\ast },s_{1}\right) +\varepsilon _{%
{\bf k}}$.

Expanding to first order in $V$ we have \cite{Note1}, 
\begin{eqnarray} \label{eq:tau3D}
\tau ^{3D}\left( {\bf k,}s,i\right)&=&-\frac{V_{h}^{s}}{\omega _{s}}%
\left\langle {\bf k}\left| \Delta \right| i\right\rangle 
\\ \nonumber
&+&\left\langle {\bf k%
}\left| V^{s}\frac{1}{\omega _{s}+\varepsilon _{{\bf k}}-h-\Sigma }\Delta
\right| i\right\rangle ,  
\end{eqnarray}
where $\Sigma $ is the self energy coming from a summation to infinite order
in $V$, and we have used the relations 
\[
\left\langle N_{B}^{\ast },s|N_{B}\right\rangle =-\frac{V_{h}^{s}}{\omega
_{s}},\;E\left( N_{B}^{\ast },s\right) =\omega _{s},
\]
\[\left\langle
N_{B}^{\ast },s\left| V\right| N_{B}\right\rangle =V^{s}\left( {\bf r}%
\right) . 
\]
The energy argument of $\Sigma $ is $\omega _{s}+\varepsilon _{{\bf k}}$.

These results are only meaningful when $\omega _{s}\neq 0$. For the moment
we take the excitation spectrum to have a gap, $\omega _{s}>\omega _{0}$,
for all $s$ except $s=0$. For $s=0$ we have 
\[
\left| \left\langle N_{B}^{\ast },0|N_{B}\right\rangle \right|
^{2}=e^{-a},\;a=\sum_{s}\left| \frac{V_{h}^{s}}{\omega _{s}}\right| ^{2}. 
\]
Also the $s\neq 0$ terms in Eq. \ref{eq:tau3D} have the $\exp \left(
-a/2\right) $ factor, when we go beyond first order in $V$.

We use plane waves for the parallel components of the fluctuation potential, 
$V^{s}\left( {\bf r}\right) =e^{i{\bf QR}}V\left( q_{z},{\bf Q},z\right) $.
Neglecting the reflected component, we similarly write the photo electron
wave function as $\psi _{{\bf k}}\left( {\bf r}\right) =e^{i{\bf KR}}\psi
_{k_{z}}^{1D}\left( {\bf K,}z\right) $. We further replace Im$\Sigma $ by $%
-i\Gamma $, and absorb $%
\mathop{\rm Re}%
\Sigma $ in $h$. The photo electron energy is $\varepsilon _{{\bf k}}=\left( 
{\bf K}^{2}+k_{z}^{2}\right) /2$. We can now simplify the last term in Eq. 
\ref{eq:tau3D} (c f Ref. \cite{Hedin98}) to become, 
\begin{eqnarray*}
^{PW}\langle {\bf K-Q}| \langle \psi _{k_{z}}^{1D}|
&&V\left( q_{z},{\bf Q},z\right)
\\ \nonumber
&\times& \frac{1}{\kappa ^{2}/2-t_{z}-\left(
V_{cryst}\left( z\right) -V_{0}\right) }\Delta| i\rangle ,
\end{eqnarray*}
where 
\begin{equation}
\frac{\kappa ^{2}}{2}=\omega _{s}+\varepsilon _{{\bf k}}-\frac{\left( {\bf %
K-Q}\right) ^{2}}{2}-V_{0}+i\Gamma _{1},  \label{kappa}
\end{equation}
\[
\Gamma _{1}=-%
\mathop{\rm Im}%
\Sigma ^{0}\left( k_{1},k_{1}^{2}/2\right) ,\;k_{1}^{2}/2=\omega
_{s}+\varepsilon _{{\bf k}}-V_{0},
\]
\[t_{z}=-\frac{1}{2}\frac{\partial ^{2}}{%
\partial z^{2}}.
\]
$^{PW}\left\langle {\bf K-Q}\right| $ is a plane wave 2D function, we have
neglected the variation of the crystal potential in the lateral directions,
the inner potential $V_{0}$ is some average of $V_{cryst}$, and $\Sigma
^{0}\left( k,\omega \right) $ is the electron gas self-energy. For the 1D
Green's function we have approximately (see Appendix C), 
\[
\left\langle z\left| \frac{1}{\kappa ^{2}/2-t_{z}-\left( V_{cryst}\left(
z\right) -V_{0}\right) }\right| z^{\prime }\right\rangle =A\psi _{\kappa
}^{<}\left( z_{<}\right) \psi _{\kappa }^{>}\left( z_{>}\right) .
\]
Here $\psi _{\kappa }^{<}$ and $\psi _{\kappa }^{>}$ are damped
Blochfunctions, $\psi _{\kappa }^{<}$ decreasing towards the surface, and $%
\psi _{\kappa }^{>}$ decreasing towards the inner of the crystal (the
crystal is on the positive half of the $z$ axis), $z_{>}=\max \left(
z,z^{\prime }\right) ,$ $z_{<}=\min \left( z,z^{\prime }\right) $, and the
coefficient $A$ is roughly $A=\left( i\kappa \right) ^{-1}$. In our
calculations we will use the simplest possible approximation $\psi _{\kappa
}^{>}\left( z\right) =\exp \left( i\kappa z\right) $, and $\psi _{\kappa
}^{<}\left( z\right) =\exp \left( -i\kappa z\right) $.

The $z$ part of the photo electron wave function is $\left[ \psi _{%
\widetilde{k}_{z}}^{>}\left( z\right) \right] ^{\ast }$, with 
\[
\frac{\widetilde{k}_{z}^{2}}{2}=\frac{k_{z}^{2}}{2}-V_{0}+i\Gamma
_{2},\;\Gamma _{2}=-%
\mathop{\rm Im}%
\Sigma ^{0}\left( k_{2},k_{2}^{2}/2\right),
\]
\begin{equation}k_{2}^{2}/2=\varepsilon _{{\bf %
k}}-V_{0}.  \label{k-vector2}
\end{equation}
We note that $\Gamma _{1}$ and $\Gamma _{2}\;$are different.

We give a few comments on the relation between the electron energy inside
the solid and outside. $V_{cryst}\left( z\right) $ is defined as $%
V_{cryst}=V_{H}+%
\mathop{\rm Re}%
\Sigma \left( \omega \right) +\phi ^{DP}$, where $\phi ^{DP}$ is the dipole
contribution to the work function $\phi $, with $\phi $ defined as negative.
For the argument $\varepsilon $ in $%
\mathop{\rm Im}%
\Sigma \left( \varepsilon \right) $ we should choose the eigenvalue in the
quasi particle equation $\left( t+V_{cryst}\left( {\bf r}\right)
-\varepsilon _{{\bf k}}\right) \psi _{{\bf k}}\left( {\bf r}\right) =0$. For
an electron gas this gives $V_{cryst}=%
\mathop{\rm Re}%
\Sigma ^{0}\left( k,\varepsilon _{k}\right) +\phi ^{DP}$ since $V_{H}=0$.
The work function is by definition $\phi =\varepsilon _{F}+%
\mathop{\rm Re}%
\Sigma ^{0}\left( k_{F},\varepsilon _{k_{F}}\right) +\phi ^{DP}$. Since $%
\mathop{\rm Re}%
\Sigma ^{0}\left( k,\varepsilon _{k}\right) $ varies fairly slowly with $k$
out to about $2k_{F}$, we can take $V_{0}=\phi -\varepsilon _{F}=-\left|
\phi \right| -\varepsilon _{F}$. When we leave the electron gas a reasonable
definition for $V_{0}$ is $V_{0}=-\left| \phi \right| -W$, where $W>0$ is
the bandwidth. The maximum kinetic energy the photo electron can have
outside the solid is, by energy conservation, $\omega _{phot}-\left| \phi
\right| $, corresponding to the energy $\omega _{phot}+W$ inside the solid.
In our calculations we have taken $\left| \phi \right| =W=0.15\;au$. Errors
in this choice have a minor effect, and the relative error decreases with
increasing photon energy. All calculations are made for a photon energy of $%
1\;au.$

We consider only forward propagation for the Green's function from the
excited layer to the surface (c f \cite{Bardy85}), and have, 
\begin{eqnarray*}
\tau ^{3D}\left( {\bf k,}s,i\right)&=&-\frac{V_{h}^{s}}{\omega _{s}}%
\left\langle {\bf k}\left| \Delta \right| i\right\rangle 
\\ \nonumber
&+&\frac{1}{i\kappa }%
\int_{0}^{\infty }\psi _{\widetilde{k}_{z}}^{>}\left( z^{\prime }\right)
V\left( q_{z},{\bf Q},z^{\prime }\right) \psi _{\kappa }^{<}\left( z^{\prime
}\right) dz^{\prime } 
\\ \nonumber
& &\times \int_{z^{\prime }}^{\infty }dz^{\prime \prime }\psi _{\kappa
}^{>}\left( z^{\prime \prime }\right) \int d{\bf R}e^{-i\left( {\bf K-Q}%
\right) {\bf R}}\Delta \left( {\bf R,}z^{\prime \prime }\right) 
\\ \nonumber
& & \ \ \ \ \ \ \ \ \ \ \ \times\psi
_{i}\left( {\bf R,}z^{\prime \prime }\right) , 
\end{eqnarray*}
or 
\begin{eqnarray*}
\tau ^{3D}\left( {\bf k,}s,i\right)&=&-\frac{V_{h}^{s}}{\omega _{s}}%
\left\langle k_{z},{\bf K}\left| \Delta \right| i\right\rangle 
\\ \nonumber
&+&\frac{1}{%
i\kappa }\int_{0}^{z_{0}}dz\psi _{\widetilde{k}_{z}}^{>}\left( z\right)
V\left( q_{z},{\bf Q},z\right) \psi _{\kappa }^{<}\left( z\right)
\\ \nonumber
& & \ \ \ \ \ \ \ \ \ \ \times
\left\langle \kappa ,{\bf K-Q}\left| \Delta \right| i\right\rangle . 
\end{eqnarray*}

If we approximate the $\psi $ functions with plane waves the dipole matrix
elements $\left\langle \widetilde{k}_{z},{\bf K}\left| \Delta \right|
i\right\rangle $ and $\left\langle \kappa ,{\bf K-Q}\left| \Delta \right|
i\right\rangle $ depend on the position of the excited layer through the
factors $\exp \left( i\widetilde{k}_{z}z_{0}\right) $ and $\exp \left(
i\kappa z_{0}\right) $. We note that the electron lifetime is $\tau
=1/\left( 2\Gamma \right) $ and from Eq. \ref{k-vector2} we have $2%
\mathop{\rm Im}%
\widetilde{k}_{z}\simeq 2\Gamma _{2}/\left| \widetilde{k}_{z}\right|
=1/\lambda $ where $\lambda =v\tau $ is the mean free path. Neglecting the
recoil momentum ${\bf Q}$ picked up by the quasi-boson $"s"$ the
entanglement between the 2D and 3D parts disappears and we have the
intuitively expected result, 
\begin{equation}
J_{{\bf k}}\left( z_{0},\omega _{phot}\right) =\int J_{{\bf k}}^{2D}\left(
z_{0},\omega \right) P_{{\bf k}}\left( z_{0},\omega _{phot}-\omega \right)
d\omega ,  \label{Jkfinal}
\end{equation}
with 
\begin{eqnarray*}
J_{{\bf k}}^{2D}\left( z_{0},\omega \right)&=&\sum_{s_{2}}\left|
\sum_{i}\left\langle N_{2D}-1,s_{2}\left| c_{i}\right| N_{2D}\right\rangle
\left\langle k,{\bf K}\left| \Delta \right| i\right\rangle \right|
^{2}
\\ \nonumber
& &\times\delta \left( \omega +E_{s_{2}}-\varepsilon _{{\bf k}}\right) , 
\end{eqnarray*}
\begin{eqnarray} \label{Pkfinal}
P_{{\bf k}}\left( z_{0},\omega \right)&=&e^{-z_{0}/\lambda -a}\left ( \delta
\left( \omega \right)
+\sum_{s}\left| -\frac{V_{h}^{s}}{\omega _{s}}
+\frac{%
e^{i\left( \kappa -\widetilde{k}_{z}\right) z_{0}}}{i\kappa }\right.\right.%
\nonumber \\
& &\times \left.\left.
\int_{0}^{z_{0}}dze^{i\left( \widetilde{k}_{z}-\kappa \right) z}V\left(
q_{z},{\bf Q},z\right) \right| ^{2}
\delta \left( \omega -\omega _{s}\right)
\right) .  
\nonumber \\
\end{eqnarray}
where $s=\left( q_{z},{\bf Q}\right) $. We have included the $s=0$ term and
the common factor $\exp \left( -a\right) $. {\it The photo current }$J_{{\bf %
k}}\left( z_{0},\omega _{phot}\right) ${\it \ thus is a convolution between
the sudden approximation 2D current }$J_{{\bf k}}^{2D}\left( z_{0},\omega
\right) ${\it , and an effective loss function }$P_{{\bf k}}\left(
z_{0},\omega \right) ${\it .}

\section{Appendix B. Dielectric response}

Dielectric response is usually treated in the random phase approximation
(RPA), and RPA\ has indeed proved extremely useful in many cases. \cite
{Savrasov96} For e g high $T_{c}$ materials RPA\ may however not be good
enough, and we will derive formal expressions without resorting to RPA.
These expressions allow us to connect the energy loss results to the
screened potentials needed to discuss photo emission. The energy loss data
are then taken from experiment. Some of our results can be found in
Griffin's classic paper \cite{Griffin88}, but not those which are crucial to
our treatment.

The response functions $\chi ^{0}$, $\chi $, and $\varepsilon ^{-1}$ are
defined from (in a schematic notation) 
\[
\rho ^{ind}=\chi ^{0}V^{tot}=\chi V^{ext},\;V^{tot}=v\rho
^{ind}+V^{ext}=\varepsilon ^{-1}V^{ext}. 
\]
This leads to the relations 
\[
\varepsilon ^{-1}=1+v\chi ,\chi =\chi ^{0}+\chi ^{0}v\chi . 
\]
Since $\rho ^{ind}$ and $V^{ext}$ are exactly defined, no approximations are
involved in the definitions of $\chi ^{0}$, $\chi $, and $\varepsilon ^{-1}$.

We now specialize to two layers per unit cell. We choose the origin of the $z
$ coordinate at the center of the cell such that we have two layers at $%
z=\pm d$. We write the response functions as, 
\[
\chi ^{0}({\bf r},{\bf r}^{\prime })=\sum_{m}\sum_{n}^{\pm
1}w(z-cm-dn)w(z^{\prime }-cm-dn)\widetilde{\chi }^{0}({\bf R}-{\bf R}%
^{\prime })
\]
\begin{eqnarray*}
\chi ({\bf r},{\bf r}^{\prime })& =&\sum_{mm^{\prime }}\sum_{nn^{\prime }}^{\pm
1}w(z-cm-dn)w(z^{\prime }-cm^{\prime }-dn^{\prime })
\\ \nonumber
& &\times\widetilde{\chi }%
_{nn^{\prime }}(m-m^{\prime };{\bf R}-{\bf R}^{\prime })
\end{eqnarray*}
We have assumed translational invariance in the layers, and that there are
no transverse excitations, i e that the electrons always stay in the lowest
transverse state $\phi _{0}\left( z\right) ,\;w\left( z\right) =\left| \phi
_{0}\left( z\right) \right| ^{2}$. We have taken the overlap between $%
w\left( z\right) $ and $w\left( z+c\right) $ as zero, and neglected
interlayer coupling in $\chi ^{0}$. This latter neglect is probably
innocent. Interlayer coupling in $\chi ^{0}$ is absent in any one electron
theory with a local potential, and thus e g in $RPA$. It is also absent in
the static ($\omega =0$) case since this case can be described by $DFT$
where the potential is local.

We Fourier transform with respect to ${\bf R}$, and separate into
contributions from different layers 
\[
\chi ^{0}(z,z^{\prime };{\bf Q})=\sum_{n}\chi _{n}^{0}(z,z^{\prime };{\bf Q}%
),
\]
\begin{equation}
\chi (z,z^{\prime };{\bf Q})=\sum_{nn^{\prime }}\chi _{nn^{\prime
}}(z,z^{\prime };{\bf Q}),  \label{chinn'}
\end{equation}
where 
\[
\chi _{n}^{0}(z,z^{\prime };{\bf Q})=\sum_{m}w(z-cm-dn)w(z^{\prime }-cm-dn)%
\widetilde{\chi }^{0}({\bf Q}),
\]
\begin{eqnarray*}
\chi _{nn^{\prime }}(z,z^{\prime };{\bf Q})&=&\sum_{mm^{\prime
}}w(z-cm-dn)w(z^{\prime }-cm^{\prime }-dn^{\prime })
\\ \nonumber
& &\times\widetilde{\chi }%
_{nn^{\prime }}(m-m^{\prime };{\bf Q}).
\end{eqnarray*}
The integral equation $\chi =\chi ^{0}+\chi ^{0}v\chi $ can be written as
(suppressing the ${\bf Q}$ variable), 
\begin{eqnarray} \label{chieq1}
\chi _{nn^{\prime }}(z,z^{\prime })&=&\chi _{n}^{0}(z,z^{\prime })\delta
_{nn^{\prime }}
\\ \nonumber
&+&\sum_{n^{\prime \prime }}\int \chi _{n}^{0}(z,z_{1})v\left(
z_{1},z_{2}\right) \chi _{n^{\prime \prime }n^{\prime }}(z_{2},z^{\prime
})dz_{1}dz_{2}.  
\end{eqnarray}
This is the same result as in Eq. 1 in Griffin's paper \cite{Griffin88}.

We Fourier transform $\chi _{n}^{0}(z,z^{\prime })$ with respect to $z$ and $%
z^{\prime }$, using discrete $q_{z}$ values and the orthonormal set $\left\{
L^{-1/2}\exp (iq_{z}z)\right\} $ 
\[
\chi _{n}^{0}(q_{z},q_{z}^{\prime })=\frac{1}{c}w\left( q_{z}\right) w\left(
q_{z}^{\prime }\right) \widetilde{\chi }^{0}e^{i\left( q_{z}-q_{z}^{\prime
}\right) dn},\; 
\]
where it is understood that $q_{z}$ and $q_{z}^{\prime }$ differ by a
reciprocal lattice vector $G$, $L$ is the length of the sample, and $c$ the
lattice constant. Similarly we have for $\chi _{nn^{\prime }}\left(
q_{z},q_{z}^{\prime }\right) $ 
\begin{equation}
\chi _{nn^{\prime }}(q_{z},q_{z}^{\prime })=\frac{1}{c}w\left( q_{z}\right)
w\left( q_{z}^{\prime }\right) \widetilde{\chi }_{nn^{\prime }}\left(
q_{z}\right) e^{iq_{z}dn}e^{-iq_{z}^{\prime }dn^{\prime }},  \label{chi1}
\end{equation}
where 
\[
\widetilde{\chi }_{nn^{\prime }}\left( q_{z}\right) =\sum_{m}\widetilde{\chi 
}_{nn^{\prime }}(m-m^{\prime })e^{iq_{z}c\left( m-m^{\prime }\right) }. 
\]
We note that $\widetilde{\chi }_{nn^{\prime }}\left( q_{z}\right) $ is a
periodic function in $q_{z}$, $\widetilde{\chi }_{nn^{\prime }}\left(
q_{z}\right) =\widetilde{\chi }_{nn^{\prime }}\left( q_{z}+G\right) $. Eqs. 
\ref{chinn'} and \ref{chi1} give 
\begin{equation}
\chi (q_{z},q_{z}^{\prime })=\frac{1}{c}w\left( q_{z}\right) w\left(
q_{z}^{\prime }\right) \sum_{nn^{\prime }}\widetilde{\chi }_{nn^{\prime
}}\left( q_{z}\right) e^{iq_{z}dn}e^{-iq_{z}^{\prime }dn^{\prime }}.
\label{chieq11}
\end{equation}
We can separate out the $w$ factors in Eq. \ref{chieq1} to obtain an
equation for $\widetilde{\chi }_{nn^{\prime }}\left( q_{z}\right) $

\begin{equation}
\widetilde{\chi }_{nn^{\prime }}\left( q_{z}\right) =\widetilde{\chi }%
_{nn^{\prime }}^{0}+\frac{1}{c}\sum_{n_{1}n_{2}}\widetilde{\chi }%
_{nn_{1}}^{0}\widetilde{V}_{n_{1}n_{2}}\left( q_{z}\right) \widetilde{\chi }%
_{n_{2}n^{\prime }}\left( q_{z}\right) ,  \label{chieq2}
\end{equation}
where $\widetilde{\chi }_{nn^{\prime }}^{0}=\widetilde{\chi }^{0}\delta
_{nn^{\prime }}$ and $\widetilde{V}$ is a 2$\times$2 
matrix periodic in $q_{z}$, 
\[
\widetilde{V}_{nn^{\prime }}\left( q_{z}\right) =\sum_{G}v\left(
q_{z}+G\right) w^{2}\left( q_{z}+G\right) e^{-i\left( q_{z}+G\right) d\left(
n-n^{\prime }\right) }. 
\]

Eq. \ref{chieq2} gives the matrix solution 
\[
\widetilde{\chi }_{nn^{\prime }}\left( q_{z}\right) =\left[ \widetilde{\chi }%
^{0}\frac{1}{1-\widetilde{V}\left( q_{z}\right) \widetilde{\chi }^{0}/c}%
\right] _{nn^{\prime }}. 
\]
We write 
\[
\widetilde{V}_{nn^{\prime }}\left( q_{z}\right) =\left( 
\begin{array}{cc}
V_{0}\left( q_{z}\right) & V_{1}\left( q_{z}\right) e^{i\phi \left(
q_{z}\right) } \\ 
V_{1}\left( q_{z}\right) e^{-i\phi \left( q_{z}\right) } & V_{0}\left(
q_{z}\right)
\end{array}
\right) , 
\]
where 
\[
V_{0}\left( q_{z}\right) =\widetilde{V}_{1,1}\left( q_{z}\right)
=\sum_{G}v\left( q_{z}+G\right) w^{2}\left( q_{z}+G\right),
\]
\[V_{1}\left(
q_{z}\right) =\left| \widetilde{V}_{1,-1}\left( q_{z}\right) \right| , 
\]
and have 
\[
\widetilde{\chi }_{nn^{\prime }}\left( q_{z}\right) =\frac{1}{2}\left( 
\begin{array}{cc}
\chi _{1}+\chi _{2} & \left( \chi _{1}-\chi _{2}\right) \exp \left( i\phi
\right) \\ 
\left( \chi _{1}-\chi _{2}\right) \exp \left( -i\phi \right) & \chi
_{1}+\chi _{2}
\end{array}
\right) , 
\]

\[
\chi _{1}=\frac{\widetilde{\chi }^{0}}{1-\left( \widetilde{\chi }%
^{0}/c\right) \left( V_{0}+V_{1}\right) },\;\chi _{2}=\frac{\widetilde{\chi }%
^{0}}{1-\left( \widetilde{\chi }^{0}/c\right) \left( V_{0}-V_{1}\right) }.
\]
Here $\chi _{1}$ and $\chi _{2}$ are functions of $q_{z},$ ${\bf Q}$ and $%
{\bf \omega }$. From Eq. \ref{chieq11} the energy loss function becomes, 
\begin{eqnarray*}
&&v\left( q_{z},{\bf Q}\right) 
\mathop{\rm Im}%
\chi \left( q_{z},q_{z};{\bf Q,\omega }\right)
\\ \nonumber
&=&\frac{v\left( q_{z},{\bf Q}%
\right) w^{2}(q_{z})}{c}%
\\ \nonumber
& &\times
\mathop{\rm Im}%
\left[ \chi _{1}+\chi _{2}+\left( \chi _{1}-\chi _{2}\right) \cos \left(
2q_{z}d+\phi \left( q_{z},{\bf Q}\right) \right) \right] ,
\end{eqnarray*}
to be compared with the screened potential $%
\mathop{\rm Im}%
W=v\left( 
\mathop{\rm Im}%
\chi \right) v$. From Eqs. \ref{chinn'} and \ref{chi1} we have 
\begin{eqnarray} \label{ImW1}
&&
\mathop{\rm Im}%
W\left( {\bf Q},\omega {\bf ;}z,z^{\prime }\right)
\\ \nonumber &=&%
\mathop{\rm Im}%
\sum_{nn^{\prime }}^{\pm 1}\frac{1}{2\pi }\int_{-\pi /c}^{\pi /c}V\left( 
{\bf Q,}q_{z},z-dn\right) \frac{\widetilde{\chi }_{nn^{\prime }}({\bf Q,}%
\omega {\bf ,}q_{z})}{c}
\\ \nonumber
& & \ \ \ \ \ \ \ \ \ \ \ \ \ \ \ \ \ \ \ \times 
V^{\ast }\left( {\bf Q,}q_{z},z^{\prime }-dn^{\prime
}\right) dq_{z},  
\end{eqnarray}
where 
\begin{equation}
V\left( Q,q_{z},z\right) =\sum_{G}v\left( q_{z}+G,Q\right)
w(q_{z}+G)e^{-i\left( q_{z}+G\right) z}.  \label{Vfluct3}
\end{equation}
There is thus no simple relation between $%
\mathop{\rm Im}%
W\left( z,z^{\prime }\right) $ and the loss function unless the non-diagonal
elements in $\widetilde{\chi }_{nn^{\prime }}(q_{z})$ can be neglected. For
typical values of $q_{z}$ and $Q$ it however turns out that $V_{1}/V_{0}$ is
0.2-0.3. Taking $\chi _{1}=\chi _{2}$ and using the symmetries $V\left(
q_{z},-z\right) =V\left( -q_{z},z\right) =V^{\ast }\left( q_{z},z\right) $
and $\chi _{n}(q_{z})=\chi _{n}(-q_{z})$ we can write Eq. \ref{ImW1} as 
\begin{eqnarray*}
\mathop{\rm Im}%
W\left( z,z^{\prime }\right)&=&\sum_{i=1}^{2}\sum_{n}^{\pm 1}\frac{1}{\pi }%
\int_{0}^{\pi /c}V_{i}\left( q_{z},z-dn\right) 
\\ \nonumber
& & \ \ \ \ \ \ \ \times
\frac{%
\mathop{\rm Im}%
\chi _{0}(q_{z})}{c}V_{i}\left( q_{z},z^{\prime }-dn\right) dq_{z},
\end{eqnarray*}
where $V_{1}$ is the real and $V_{2}$ the imaginary part of $V\left(
q_{z},z\right) $ and $\chi _{0}=\left( \chi _{1}+\chi _{2}\right) /2$. The
real and imaginary parts turn out to give equal contributions to $%
\mathop{\rm Im}%
W$.

So far we have results for a set of coupled layers sitting in vacuum. We can
take account of the embedding electrons (the 3D bulk excitations in our
parametrization) by using 
\begin{eqnarray} \label{epsbulk}
\chi ^{0}\left( z,z^{\prime };{\bf Q}\right)&=&\chi _{b}^{0}\left(
z-z^{\prime };{\bf Q}\right) 
\\ \nonumber
&+&\sum_{m}w(z-cm)w(z^{\prime }-cm)\widetilde{%
\chi }_{2}^{0}\left( {\bf Q}\right) . 
\end{eqnarray}
This leads to a 3D bulk contribution in $%
\mathop{\rm Im}%
W\left( z,z^{\prime }\right) $, and to screening of the 2D susceptibility $%
\chi $. The same screening however appears also in the loss function, so we
can forget about it in our problem. We note that the bulk screened potential
can be anisotropic since $\chi _{b}^{0}\left( q_{z},{\bf Q}\right) $ can
depend on both $q_{z}$ and ${\bf Q}$, and not only on $q^{2}=q_{z}^{2}+{\bf Q%
}^{2}$.

We derived the relation between $\chi $ and $\chi ^{0}$ by solving the
integral equation $\chi =\chi ^{0}+\chi ^{0}v\chi $. This equation can be
written as $\chi =\chi _{2D}+\chi _{2D}v_{3D}\chi $, where $\chi _{2D}=\chi
^{0}+\chi ^{0}v_{2D}\chi _{2D}$. Here $v_{3D}$ contains no intralayer parts,
while $v_{2D}$ only has intralayer contributions. Since $\chi _{2D}$ is
available from many sophisticated theoretical calculations, it is
interesting to have the relation between $\chi ^{0}=\chi _{2D}^{0}$ and $%
\chi _{2D}$. We write 
\[
\chi _{2D}\left( z,z^{\prime }\right) =w\left( z\right) w\left( z^{\prime
}\right) \widetilde{\chi }_{2D},
\]
\[\chi _{2D}^{0}\left( z,z^{\prime }\right)
=w\left( z\right) w\left( z^{\prime }\right) \widetilde{\chi }_{2D}^{0}. 
\]
The equation $\chi _{2D}=\chi ^{0}+\chi ^{0}v_{2D}\chi _{2D}$ gives, $%
\widetilde{\chi }_{2D}=\widetilde{\chi }_{2D}^{0}+\widetilde{\chi }%
_{2D}^{0}W_{00}\widetilde{\chi }_{2D},$ where $W_{00}=\int w\left(
z_{1}\right) v\left( z_{1}-z_{2};{\bf Q}\right) w\left( z_{2}\right)
dz_{1}dz_{2}.$ The desired relation is, 
\[
\widetilde{\chi }_{2D}^{0}=\frac{\widetilde{\chi }_{2D}}{1+W_{00}\widetilde{%
\chi }_{2D}}. 
\]

\section{Appendix C. On Green's functions when the potential is complex}

Green's function theory is usually developed using a\ real potential. Here
we will shortly summarize the changes from having a complex potential. With
a constant complex potential the equation ($%
\mathop{\rm Im}%
\kappa >0$),

\[
\frac{1}{2}\left( \kappa ^{2}+\frac{d^{2}}{dz^{2}}\right) G(z,z^{\prime
};\kappa )=\delta (z-z^{\prime }), 
\]
has the solution (as is easily verified by direct substitution), 
\[
G(z,z^{\prime };\kappa )=\frac{1}{i\kappa }e^{i\kappa \left| z-z^{\prime
}\right| }. 
\]
With $\kappa $ a function of $z$ the solution has the form, 
\[
G\left( z,z^{\prime };\kappa \right) =wg_{-}\left( z_{<}\right) g_{+}\left(
z_{>}\right) , 
\]
where $w$ is a constant (c f e g Arfken, \cite{Arfken}), 
\[
w=\frac{2}{g_{-}\left( z\right) g_{+}^{\prime }\left( z\right)
-g_{-}^{\prime }\left( z\right) g_{+}\left( z\right) }. 
\]
The boundary conditions are $g_{-}\left( z\right) \rightarrow 0$ for $%
z\rightarrow -\infty $ and $g_{+}\left( z\right) \rightarrow 0$ for $%
z\rightarrow \infty $.

In a slightly more general situation

\[
\kappa \left( z\right) =\left\{ 
\begin{array}{c}
\kappa _{1},\;z<0 \\ 
\kappa _{2},\;z>0
\end{array}
\right. ,\;%
\mathop{\rm Im}%
\kappa _{i}>0, 
\]
we have 
\[
g_{-}\left( z\right) =e^{-i\kappa _{1}z}\theta \left( -z\right) +\left(
a_{-}e^{i\kappa _{2}z}+b_{-}e^{-i\kappa _{2}z}\right) \theta \left( z\right)
, 
\]
\[
g_{+}\left( z\right) =e^{i\kappa _{2}z}\theta \left( z\right) +\left(
a_{+}e^{i\kappa _{1}z}+b_{+}e^{-i\kappa _{1}z}\right) \theta \left(
-z\right) , 
\]
\[
a_{-}=\frac{1}{2}\left( 1-\frac{\kappa _{1}}{\kappa _{2}}\right) ,\;b_{-}=%
\frac{1}{2}\left( 1+\frac{\kappa _{1}}{\kappa _{2}}\right) , 
\]
\[
a_{+}=\frac{1}{2}\left( 1+\frac{\kappa _{2}}{\kappa _{1}}\right) ,\;b_{+}=%
\frac{1}{2}\left( 1-\frac{\kappa _{2}}{\kappa _{1}}\right) , 
\]
\[
w=\frac{2}{i\left( \kappa _{1}+\kappa _{2}\right) }. 
\]
For high energies $\kappa _{1}\simeq \kappa _{2}$ and the results reduce to
those of the first model with $\kappa $=constant.

Finally we consider a model with,

\[
\kappa \left( z\right) =\left\{
\begin{array}{c}
\kappa _{1},\;z<0 \\ 
\kappa _{2}\left( z\right) ,\;z>0
\end{array}
\right. ,\;%
\mathop{\rm Im}%
\kappa _{i}>0, 
\]
\[
g_{-}\left( z\right) =e^{-i\kappa _{1}z}\theta \left( -z\right) +\left(
a_{-}\psi _{\kappa }^{>}\left( z\right) +b_{-}\psi _{\kappa }^{<}\left(
z\right) \right) \theta \left( z\right) , 
\]
\[
g_{+}\left( z\right) =\psi _{\kappa }^{>}\left( z\right) \theta \left(
z\right) +\left( a_{+}e^{i\kappa _{1}z}+b_{+}e^{-i\kappa _{1}z}\right)
\theta \left( -z\right) . 
\]
To see the difference between $\psi _{\kappa }^{<}\left( z\right) $ and $%
\psi _{\kappa }^{>}\left( z\right) $ we use the WKB\ approximation for the
case $z>0$, 
\[
\psi _{\kappa }^{>}\left( z\right) \sim e^{i\int_{0}^{z}\kappa _{2}\left(
z^{\prime }\right) dz^{\prime }},\;\psi _{\kappa }^{<}\left( z\right) \sim
e^{-i\int_{0}^{z}\kappa _{2}\left( z^{\prime }\right) dz^{\prime }}. 
\]
Since $\kappa _{2}\left( z\right) $ is complex, the two solutions are
basically different, not just complex conjugates. For higher energies and $\
z>0$ we have $g_{-}\left( z\right) \simeq \psi _{\kappa }^{<}\left( z\right)
,\;g_{+}\left( z\right) =\psi _{\kappa }^{>}\left( z\right) $.

\end{multicols}

\end{document}